\documentclass{aa}  
\usepackage{graphicx}
\usepackage{txfonts}
\usepackage{supertabular}
\begin{document}
    \title{The diversity of methanol maser morphologies from VLBI observations
\thanks{Tables 1-3 and 6, Figures 3 and 6 are only available in electronic form via
http://www.aanda.org} 
}
   \author{A. Bartkiewicz
          \inst{1},
          M. Szymczak
          \inst{1},
          H.J. van Langevelde
          \inst{2,3},
          A.M.S. Richards\inst{4},
          \and
          Y.M. Pihlstr\"om\inst{5,6}
          }

   \institute{Toru\'n Centre for Astronomy, Nicolaus Copernicus
          University, Gagarina 11, 87-100 Toru\'n, Poland\\
          \email{[annan;msz]@astro.uni.torun.pl}
\and      Joint Institute for VLBI in Europe, Postbus 2, 7990 AA
          Dwingeloo, The Netherlands\\
          \email{langevelde@jive.nl}
\and      Sterrewacht Leiden, Postbus 9513, 2300 RA Leiden, The Netherlands
\and      Jodrell Bank Centre for Astrophysics, Alan Turing Building, University of Manchester, 
          M13 9PL, UK\\
          \email{a.m.s.richards@manchester.ac.uk}
\and      Department of Physics and Astronomy, MSC07 4220, University of New Mexico, Albuquerque, 
          NM 87131, USA\\
          \email{ylva@unm.edu}
\and      National Radio Astronomy Observatory, 1003 Lopezville Road, Socorro, NM 87801, USA
          }

   \date{Received; accepted }

\authorrunning{Bartkiewicz et al.}
\titlerunning{The diversity of methanol masers morphologies from VLBI}

% \abstract{}{}{}{}{} 
% 5 {} token are mandatory
 
  \abstract % context heading (optional) 
{The 6.7\,GHz methanol maser marks an early stage of high--mass star 
formation, but the origin of this maser is currently a matter of debate. 
In particular it is unclear whether the maser emission arises in discs, 
outflows or behind shocks running into rotating molecular clouds.}  
{We investigate which structures the methanol masers trace in
the environment of high-mass protostar candidates by observing a
homogenous sample of methanol masers selected from Torun surveys. 
We also probed their origins by looking for associated H\,{\small II} regions 
and IR emission.} 
{We selected 30 methanol sources with improved position accuracies 
achieved using MERLIN and another 3 from the literature. We imaged 
31 of these using the European VLBI Network's expanded array of 
telescopes with 5-cm (6-GHz) receivers. We used the VLA to search 
for 8.4\,GHz radio continuum counterparts and inspected {\emph Spitzer} 
GLIMPSE data at 3.6--8\,$\mu$m from the archive.} 
{High angular resolution images allowed us to analyze the morphology
and kinematics of the methanol masers in great detail and verify
their association with radio continuum and mid-infrared emission. A
new class of\, "ring-like" methanol masers in star--forming regions
appeared to be suprisingly common, 29\% of the sample.}  
{The new morphology strongly suggests that methanol masers originate in the
disc or torus around a proto- or a young massive star. However, the
maser kinematics  indicate the strong influence of outflow
or infall. This suggests that they form at the interface between the
disc/torus and a flow. This is also strongly supported
by {\emph Spitzer} results because the majority of the masers coincide with
4.5\,$\mu$m emission to within less than 1\arcsec. Only four masers are
associated with the central parts of UC H\,{\small II} regions. This
implies that 6.7\,GHz methanol maser emission occurs before H\,{\small II}
region observable at cm wavelengths is formed.}  
%{E}


   \maketitle
%
%________________________________________________________________

\section{Introduction}
Methanol masers are commonly assumed to be associated with the
environments of high-mass protostars, which provide the conditions required for
methanol first to form on grains and then to be sublimated off, and
finally to excite the maser transitions (Dartois et al. \cite{dartois99}; 
Cragg et al. \cite{cragg02}).  Methanol maser emission at
6668.519\,MHz is one of the strongest and most widespread (Menten
\cite{menten91}) of the the first observable manifestations of a newly
formed high-mass star. Emission towards the archetypical source W3(OH)
is characterized with a brightness temperature of up to
$\sim$$3\times10^{12}$\,K from a spot of intrinsic size of
$\sim$0\farcs0014 (Menten et al. \cite{menten92}). The gas environment
of even distant ($>$5\,kpc) high-mass stars can thus be probed on
scales as small as $\sim$5\,AU when observed with milliarcsecond
(mas) resolution using Very Long Baseline Interferometry (VLBI).

All major methanol targeted surveys of high angular resolution taken
to date are summarized in Table \ref{table:1}. This indicates the
diversity of the sample selections and observing parameters, which
might have affected the data interpretation.
 
The first observations of the 6.7\,GHz maser line at arcsec
resolution, using the ATCA, concentrated on the brightest sources
(Norris et al.\,\cite{norris93}), followed by more extensive surveys
(Phillips et al.\,\cite{phillips98}; Walsh et
al.\,\cite{walsh98}). 
The relative positions of individual maser spots were determined with
$\sim$0\farcs05 accuracy and the distribution of bright
($>$0.5\,Jy\,beam$^{-1}$) maser spots was resolved for the majority of
targets. Various morphological structures, such as simple, linear,
curved, complex and double, were found. The linear sizes varied between
190 and 5600\,AU.  Norris et al. (\cite{norris93}) found that in 10 out
of the 15 sources imaged the masers are located along lines or arcs of
which five sources show a clear velocity gradient along the line. They
proposed that the linear structures with velocity gradients are produced by 
the masers residing in rotating discs seen edge-on. Phillips et
al. (\cite{phillips98}) increased the sample of masers studied with the
ATCA to 45 objects, finding that 17 of them show morphologies and
monotonic velocity gradients consistent with the circumstellar disc
hypothesis.  Assuming a Keplerian disc, the enclosed masses range from
1 to 75 M$_{\sun}$.  In a sample of 97 sources Walsh et
al. (\cite{walsh98}) found 36 masers with some linear structure but
this was clearly-defined for only 9 sources.  Therefore, the Keplerian
disc hypothesis accounts for only a small proportion of their sources;
most maser sites do not exhibit a systematic velocity gradient. They
suggested that the masers form rather behind shock fronts.

The methanol masers are often not associated with detectable continuum
emission at centimeter wavelengths. Twenty-five of sources in the sample of
Phillips et al.  (\cite{phillips98}) are associated with an
ultra-compact H\,{\small II} (UC\,H\,{\small II}) region wherein the
methanol masers are slightly offset from the peak continuum
emission. They argued that the methanol sources without an H\,{\small
II} region are possibly associated with less massive stars than  those
with coincident radio continuum emission.  Walsh et
al. (\cite{walsh98}) also found that most of their maser sources are
not associated with radio continuum brighter than $\sim$1\,mJy,
implying that the phase of methanol maser occurs before an observable
UC\,H\,{\small II} region is formed. This suggestion was confirmed for
another sample of high-mass protostellar candidates (Beuther et
al.\,\cite{beuther02}).

The detailed spatial structure of the methanol maser emission in these
sources should provide further clues to their origin. To date, only a
few observations at mas resolution have been published. Minier et
al. (\cite{minier00}) observed 14 bright sources with the EVN. In 10
targets they found elongated structures with linear velocity gradients, 
which can be interpreted in terms of a circumstellar edge-on disc
model. However, the estimates of central mass with this model for
all but one source seemed to be far lower than expected for a
high-mass star. Minier et al. (\cite{minier00}) suggested that this
could be because the detectable masers delineate only part of the
full diameters of the discs. They also proposed other models, such as
accelerating outflows and shock fronts. Dodson et
al. (\cite{dodson04}) used the LBA to image five maser sites with 
linear morphologies at arcsecond resolutions. Their milliarcsecond
resolution data were interpreted using a model of an externally generated
planar shock propagating through a rotating dense molecular clump or
star-forming core.

Van der Walt et al. (\cite{vanderwalt07}) argued that the
model of Dodson et al. is inconsistent with the observed kinematic
properties of the masers. They concluded that the
observed rest frame distribution of maser velocities can be reproduced well
with a simple Keplerian-like disc model. Source
NGC\,7538 IRS\,1 is understood to be a good example of an edge-on
Keplerian disc (Minier et al. \cite{minier00}; Pestalozzi et
al. \cite{pestalozzi04}). However, high angular resolution
mid-infrared (MIR) data were used to demonstrate that the outflow scenario is also
plausible since the maser is not oriented perpendicular to the outflow as
expected (De Buizer \& Minier \cite{debuizer05}). The kinematic and
spatial distribution of the 12\,GHz methanol masers in W3(OH) were 
successfully fitted by a model of a conical bipolar outflow (Moscadelli
et al. \cite{moscadelli02}).

The initial methanol imaging surveys were mostly of relatively low
(arcsec) resolution, whilst very long baseline
interferometer (VLBI) studies probably missed fainter emission, such as
from the edges or the far side of putative discs. For the first time,  
we have studied a large sample, detecting 31 sources at mas resolution and 
$\sim$10\,mJy sensitivity, with sufficient astrometric precision to
complete robust identifications. This enables us to test the
competing hypotheses for the origins of methanol 6.7\,GHz maser
emission, namely circumstellar discs, outflows, or propagating shock 
fronts. In
this paper, we present EVN\footnote{The European VLBI Network}
observations of the methanol line and VLA observations of continuum
emission for a homogeneous sample of the methanol masers discovered in
the Torun untargeted survey (Szymczak et al. \cite{szymczak00},
\cite{szymczak02}). The preliminary results of methanol observations
were partly published in Bartkiewicz et al. (\cite{bartkiewicz04},
\cite{bartkiewicz06}, \cite{bartkiewicz09}). As part of this survey 
the discovery of a ring structure in G23.657$-$00.127 was reported by 
Bartkiewicz et al. (\cite{bartkiewicz05}). Our observations have enabled us
to detect a wide diversity of methanol maser geometries and demonstrate 
for the first time that in a large fraction of sources the distribution 
of the spots is ring-like.

\onltab{1}{                                                                   
\begin{table*}
\caption{Summary of previous high angular resolution studies of 6.7\,GHz methanol masers.}
\label{table:1}
\begin{tabular}{lcccccc}
\hline\hline\\\
Survey & Telescope & Spectral resolution & Angular resolution &  1$\sigma_{\mathrm{rms}}$ & Number of targets & Median peak flux 
density \\ 
       &           & (km\,s$^{-1}$)       &  (mas)             & (mJy\,beam$^{-1}$) & &  (Jy) \\
\hline
Norris et al. 1993  & ATCA& 0.3 & 1500    & $\sim$500 & 15 & 475 \\
Phillips et al. 1998& ATCA& 0.35& 1500   & $\sim$50& 33 & 41 \\
Walsh et al. 1998   & ATCA& 0.18& 1500   &       300 & 97 & 23 \\
Minier et al. 2000  & EVN & 0.04& few    &        $-^*$  & 14 & 272 \\
Dodson et al. 2004  & LBA & 0.2 & 5.3    &         2 &  5 & 148 \\
This paper          & EVN & 0.09; 0.18& $\sim6\times14$ &   3$-$12 &  31 &  3.6 \\
\hline\hline
\multicolumn{7}{l}{$^*$ the value not given}\\
\end{tabular}
\end{table*}
}

\section{Observations and data reduction}
\subsection{Sample selection}
The sources were selected from two previous samples obtained using the
Torun 32\,m antenna: the blind survey of the 6.7\,GHz methanol maser
line (Szymczak et al.\,\cite{szymczak02}) and the methanol survey of
IRAS-selected objects (Szymczak et al.\, \cite{szymczak00}). The
untargeted flux-limited (3$\sigma_{\mathrm{rms}}$$\simeq$1.6\,Jy) complete survey of
the Galactic plane region $20^\circ\leq l\leq 40^\circ$ and $\vert
b\vert\leq 0\hbox{$.\!\!^\circ$ }52$ enabled the detection of 100
sources of which 26 were new. The same field includes 22 sources
discovered in the earlier survey of IRAS-selected objects. These 48 objects 
were chosen as a sample for detailed studies. 
We note that the mean single-dish methanol maser flux density
of these 48 sources is 16\,Jy, a factor of 2
lower than that of the rest sources in the original samples. 
This may have introduced a selection effect for masers that are
more distant, have intrinsically weaker maser emission, or are less
aligned with the line of sight.

Depending on the maser flux densities the source coordinates
obtained using the 32\,m dish are accurate to within 25$-$70\arcsec.
 We undertook astrometric
measurements using the first two MERLIN\footnote{The Multi-Element
Radio Linked Interferometer Network} antennas to be equipped with
6.7-GHz receivers (Mark\,II and Cambridge). These single baseline
observations detected 30 of the 48 objects, providing positions with
sub-arcsecond accuracy (see Sect.~2.2). 
We included three additional objects in the same region that had not been 
detected in the Torun surveys, for which accurate positions are reported in the
literature. These are: G22.357$+$00.066 (Walsh et al. \cite{walsh98}),
G25.411$+$00.105, and G32.992$+$00.034 (Beuther et
al. \cite{beuther02}). The total sample selected for VLBI
observations comprised 33 sources in the Galactic within the region defined by
$21.4^\circ\leq l\leq 39.1^\circ$ and $-0.38^\circ\leq b\leq
0.56^\circ$ (Table \ref{table:2}).

\onltab{2}{
\begin{table}
\caption{Sample of methanol masers observed with the EVN. The names
are the Galactic coordinates of the brightest spot of each target
obtained in post-processing EVN data. The dates of each observing run
are listed in Table \ref{table:4}. The phase-calibrator names and
angular separations from the targets are given.}
\label{table:2}      
\centering          
\begin{tabular}{cccc}     
\hline\hline       
\\
Source&  Observing & Phase-calibrator& Separation \\
Gll.lll$\pm$bb.bbb& run      &                  &  (\fdg) \\
\hline                    
\\
 G21.407$-$00.254 & 3a & J1825$-$0737 & 3.1 \\
 G22.335$-$00.155 & 3a & J1825$-$0737 & 2.5 \\
 G22.357$+$00.066$^1$ & 4a & J1825$-$0737 & 2.3 \\
 G23.207$-$00.377 & 2  & J1825$-$0737 & 2.6 \\
 G23.389$+$00.185 & 2  & J1825$-$0737 & 2.1 \\
 G23.657$-$00.127 & 2  & J1825$-$0737 & 2.4 \\
                  & 3a & J1825$-$0737 & 2.4 \\
                  & 4a & J1825$-$0737 & 2.4 \\                 
 G23.707$-$00.198 & 2  & J1825$-$0737 & 2.5 \\
                  & 3a & J1825$-$0737 & 2.5 \\
 G23.966$-$00.109 & 3a & J1825$-$0737 & 2.5 \\
 G24.148$-$00.009 & 3a & J1825$-$0737 & 2.4 \\
 G24.541$+$00.312 & 2  & J1825$-$0737 & 2.4 \\
 G24.634$-$00.324 & 4a & J1825$-$0737 & 2.9 \\
 G25.411$+$00.105$^2$ & 4a & J1825$-$0737 & 3.1 \\
 G26.598$-$00.024 & 4a & J1825$-$0737 & 4.1 \\
 G27.221$+$00.136 & 2  & J1825$-$0737 & 4.5 \\
 G28.817$+$00.365 & 4b & J1834$-$0301 & 2.2 \\
 G30.318$+$00.070 & 4b & J1834$-$0301 & 3.2 \\
 G30.400$-$00.296 & 4b & J1834$-$0301 & 3.5 \\
 G31.047$+$00.356 & 4b & J1834$-$0301 & 3.5 \\
 G31.156$+$00.045 & 4b & J1834$-$0301 & 3.8 \\
 G31.581$+$00.077 & 4b & J1834$-$0301 & 4.1 \\
 G32.992$+$00.034$^2$ & 4c & J1907$+$0127 & 4.2 \\
 G33.641$-$00.228 & 1  & J1907$+$0127 & 3.5 \\
 G33.980$-$00.019 & 4c & J1907$+$0127 & 3.5 \\
 G34.751$-$00.093 & 4c & J1907$+$0127 & 3.0 \\
 G35.793$-$00.175 & 1  & J1907$+$0127 & 2.7 \\
 G36.115$+$00.552 & 1  & J1907$+$0127 & 3.4 \\
                  & 4c & J1907$+$0127 & 3.0 \\
 G36.705$+$00.096 & 4c & J1907$+$0127 & 3.1 \\
 G37.030$-$00.039 & 3b & J1856$+$0610 & 2.6 \\
 G37.479$-$00.105 & 3b & J1856$+$0610 & 2.4 \\
 G37.598$+$00.425 & 3b & J1856$+$0610 & 1.9 \\
 G38.038$-$00.300 & 3b & J1856$+$0610 & 2.2 \\
 G38.203$-$00.067 & 3b & J1856$+$0610 & 2.0 \\
 G39.100$+$00.491 & 3b & J1856$+$0610 & 1.2 \\
\\
\hline\hline
\multicolumn{4}{l}{$^1$ source from Walsh et al. (\cite{walsh98})}\\
\multicolumn{4}{l}{$^2$ sources from Beuther et al. (\cite{beuther02})}\\
\end{tabular}                                                                                
\end{table}
}

\subsection{MERLIN astrometry}
The MERLIN observations at a rest frame frequency of 6668.519\,MHz were
carried out during observing runs between 2002 May and June and 2003 March
and May. The typical on-source observing time was about 1\,hr for each
target, and frequent observations of nearby phase reference sources
and other calibrators were completed. Standard single-baseline data reduction
procedures were applied (Diamond et al. \cite{diamond03}) using AIPS
(the Astronomical Image Processing System). We searched for emission
from each target in its vector-averaged spectrum by shifting the phase
center from $-$200\arcsec\, to $+$200\arcsec\, in right ascension and
from $-$500\arcsec\, to $+$500\arcsec\, in declination (at 1\arcsec\,
intervals) to locate the position giving the maximum intensity for the
main maser feature. We simultaneously inspected the phase, which
should be close to 0\degr\, at this position in the spectral channels
containing the main feature. Finally, we produced a large
(40\arcsec$\times$40\arcsec) dirty map of the main feature for the
channel of the highest spectral signal-to-noise ratio, centered on
the estimated position.  The brightest spot was then assumed to be as the
maser position. A typical beam was 200$\times$20\,mas$^2$ at a
position angle of 20\degr. Because of the very poor {\it uv}$-$coverage, 
we were unable to derive the maser structures.

Methanol maser emission was detected towards 30 of the 48 sources
observed, giving absolute positions of sufficient accuracy for
follow-up EVN observations. The MERLIN single-baseline astrometric
accuracy was between 0\farcs3 and 1\arcsec\, in most cases, depending on the
source brightness. However, the absolute position uncertainty of
sources at $|$Dec$|$$<$3\fdg5 increased to 5--10\arcsec. 
The mean differences between the coordinates obtained
using a single dish and using the MERLIN single baseline were
30\arcsec$\pm$6\arcsec\, and 20\arcsec$\pm$4\arcsec\, in right
ascension and declination, respectively. No emission was detected
towards the remaining 18 sources above a sensitvity limit of 0.3\,Jy
(Table \ref{table:3}). The possible causes of non-detection are:
variability, large errors in the single dish positions, interference
(for a few targets), or extended emission, resolved out by the
interferometer.

\onltab{3}{
\begin{table}
\caption{Targets not detected with the MERLIN single baseline. The
source names, coordinates, and peak flux velocities are taken from
Szymczak et al. (\cite{szymczak02}).}
\label{table:3}      
\centering          
\begin{tabular}{cccc}     
\hline\hline       
\\
Source&  RA, Dec & V$_{\rm p}$ \\
Gll.ll$\pm$bb.bb& (J2000) & (km s$^{-1}$)\\
\hline                    
\\
G21.57$-$00.03 & 18 30 36.5, $-$10 06 43 &$+$117 \\
G22.05$+$00.22 & 18 30 35.7, $-$09 34 26 &$+$54  \\
G24.93$+$00.08 & 18 36 29.2, $-$07 05 05 &$+$53  \\
G26.65$+$00.02 & 18 39 51.8, $-$05 34 52 &$+$107 \\
G27.21$+$00.26 & 18 40 03.8, $-$04 58 09 &$+$9   \\
G27.78$+$00.07 & 18 41 47.5, $-$04 33 11 &$+$112 \\
G28.02$-$00.44 & 18 44 02.1, $-$04 34 14 &$+$17  \\
G28.40$+$00.07 & 18 42 54.5, $-$04 00 04 &$+$69  \\
G28.53$+$00.12 & 18 42 57.7, $-$03 51 59 &$+$25  \\
G28.69$+$00.41 & 18 42 13.6, $-$03 35 07 &$+$94  \\
G28.85$+$00.50 & 18 42 12.8, $-$03 24 26 &$+$83  \\
G29.31$-$00.15 & 18 45 23.1, $-$03 17 23 &$+$48  \\
G33.74$-$00.15 & 18 53 26.9, $+$00 39 01 &$+$54  \\
G33.86$+$00.01 & 18 53 05.2, $+$00 49 36 &$+$67  \\
G34.10$+$00.01 & 18 53 31.9, $+$01 02 26 &$+$56  \\
G37.53$-$00.11 & 19 00 14.4, $+$04 02 35 &$+$50  \\
G38.12$-$00.24 & 19 01 47.6, $+$04 30 32 &$+$70  \\
G38.26$-$00.08 & 19 01 28.7, $+$04 42 02 &$+$16  \\
\\
\hline\hline
\end{tabular}                                                                                
\end{table}
}

\subsection{EVN observations}
The EVN observations of 33 targets in the 6.7\,GHz methanol maser line 
were carried out in seven observing runs between 2003 and 2007 (projects
EN001, EN003, EB031, EB034). The observing parameters are summarized in
Table \ref{table:4} including the date, duration of each run,
working antennas, cycle time between the maser and phase-calibrator,
spectral resolution, typical synthesized beam size, and 1$\sigma_{\mathrm{rms}}$ noise
level in a spectral channel.

\begin{table*}
\caption{Details of EVN observations.}
\label{table:4}      
\centering          
\begin{tabular}{cccccccc}
\hline\hline
\\
Observing & Date & Duration& Telescopes$^*$ & Cycle time & Spectral channel  & Synthesized & Rms noise\\
run       &      &         &               &             & separation & beam($\alpha\times\delta$;PA)& per channel\\
          &      & (hr)    &               & (min)       &  (km\,s$^{-1}$)& (mas$\times$mas;\degr)& (mJy\,beam$^{-1}$)\\
\hline  
\\                  
1         & 2003 Jun 08& 12& CmJbEfOn          & 6.00$+$4.00& 0.09 &6$\times$16; $-$1& 10\\
2         & 2004 Nov 11& 12& CmDaEfMcNtOnTrWb  & 3.75$+$1.75& 0.09 &6$\times$16; $+$7& 4\\
3a        & 2006 Feb 22& 10& CmJbEfHhMcNtOnWb  & 3.75$+$1.75& 0.09 &5$\times$15;$+$31& 10\\
3b        & 2006 Feb 23& 10& CmJbEfHhMcOnWb    & 3.75$+$1.75& 0.09 &6$\times$14;$+$34& 7\\
4a        & 2007 Jun 13& 10& CmJbEfHhMcNtOnTrWb& 3.25$+$1.75& 0.18 &6$\times$12;$+$20& 4\\
4b        & 2007 Jun 14& 10& CmJbEfHhMcNtOnTrWb& 3.25$+$1.75& 0.18 &6$\times$11;$+$25& 4\\
4c        & 2007 Jun 15& 10& CmJbEfMcNtOnTrWb  & 3.25$+$1.75& 0.09 &6$\times$13;$+$35& 6\\
\hline\hline                   
\multicolumn{8}{l}{$^*$ Cm--Cambridge, Da--Darnhall, Jb--Jodrell Bank,
Ef--Effelsberg, Hh--Hartebeesthoek, Mc--Medicina, Nt--Noto, On--Onsala,
Tr--Toru\'n,}\\
\multicolumn{8}{l}{Wb--Westerbork}\\
\end{tabular}                                                                                
\end{table*}

Each observing run included scans of 3C345, which was adopted as a 
bandpass, delay and rate
calibrator. Five or six sources were, typically, observed in each
run, selected to be within a few degrees of each other in projection
on the sky and of similar maser emission velocities. A
phase-referencing scheme was applied in which a nearby, sufficiently bright
phase-calibrator for each session was selected from the VLBA
list. Details are given in Table \ref{table:2}. We used a spectral
bandwidth of 2\,MHz yielding a velocity coverage of
$\sim$100\,km\,s$^{-1}$. In all sessions, the Mk\,V recording system
was used with the exception of the first epoch when data were recorded
on tapes (Mk\,IV system). The data were correlated with the Mk\,IV Data
Processor operated by JIVE with 1024 spectral channels.  In 4a and 4b
runs only, when all nine antennas were operating, data were correlated with
512 spectral channels in two passes i.e., separately for LHC and RHC
polarization, because of the correlator limitations. Left- and right-hand
circular polarization data were averaged to increase the
signal-to-noise ratio.

\subsubsection{Calibration and imaging}                                                         
The data calibration and reduction were carried out in AIPS, employing
standard procedures for spectral line observations. First, the
amplitude was calibrated using measured antenna gain curves and system
temperatures. In the second step, the parallactic angle corrections
were added. The Effelsberg antenna was used as a reference when calibrating
the data from all sessions. The instrumental delays for
each antenna were determined using 3C345. Rates and delays were
calibrated using phase-calibrator and 3C345 observations. The
phase-calibrator was mapped and a few iterations of self-calibration
were completed, gradually shortening the time interval from 120\,min to
1\,min. Flux densities of 240, 202, 80, and 237\,mJy were obtained for
phase-calibrators J1825$-$0737, J1834$-$0301, J1856$+$0610, and
J1907$+$0127, respectively.  The maser data were corrected for the
effects of the Earth's rotation and its motion within the Solar System
and towards the LSR.  After applying all corrections from the
calibration sources, we compiled preliminary maps of the channel
containing the brightest and most compact peak. We then used the clean
components of that map, if possible, as the starting model for further
rounds of self-calibration. In two cases, high quality images could only be
obtained after the first round of self-calibration using a default model
at the pointing position.

We searched for emission using large (2\arcsec$\times$2\arcsec) maps
over the entire band.  We then created naturally-weighted
0.5\arcsec$\times$0.5\arcsec\, cleaned images to use in analyzing maser
properties. The beam sizes for each data set are listed in Table
\ref{table:4}. The pixel separation was 1\,mas. The rms noise levels
in line-free channels were typically between 4 and 10\,mJy beam$^{-1}$
depending on the run. The positions of all maser spots (above
5$\sigma_{\mathrm{rms}}$) in each individual channel map were
determined by fitting two-dimensional Gaussian components.  The formal
fitting errors were, typically, 0.01--0.15\,mas in right ascension and
0.02--0.5\,mas in declination, depending on the source strength and
structure.  

The astrometric accuracy for the 29 sources with
phase-referenced maps is limited by four factors. Firstly, the
phase-reference source positions have an accuracy of $<$1.5\,mas.  Secondly,
the antenna positions have an accuracy of $\sim$1\,cm, corresponding to 
an uncertainty of $\sim$1\,mas in RA and 2--3\,mas in Dec. Thirdly, the separations
between the targets and phase reference sources were $\la$4\fdg5,
which translate into a potential phase solution transfer error equivalent to 2\,mas
in RA and 4--5\,mas in Dec. Fourthly, the position uncertainty due to
noise is given by (beamsize)/(signal-to-noise ratio), which is $\ll$1 mas
for all our reference features. This infers a total astrometric
uncertainty of 3\,mas in RA and 6\,mas in Dec. For the remaining four
sources, we were unable to improve on the original MERLIN positions.

\subsection{VLA continuum observations}
In order to investigate the presence, position, and distribution of radio 
continuum emission associated with the 6.7\,GHz methanol maser emission, we used
the VLA at 8.4\,GHz in A configuration (the project AB1250).
Data were taken on 2007 August 18 for 12\,hrs in a standard VLA continuum mode
towards 30 sources in the sample. We did not observe the three sources
that had not been included in the Torun surveys. 
We used 3C286 as a flux calibrator and two
phase-calibrators, 18517$+$0355 and 18323$-$1035, from a standard VLA list. 
To increase signal-to-noise ratio,  
we employed the fast switching mode with a cycle time between the phase-calibrator 
and the target of 50\,s$+$250\,s. This sequence lasted for 20\,min for each target. 

The data reduction was carried out following to the standard recipies
from AIPS Cookbook Appendix A (NRAO 2007). The amplitude and phases of
3C286 were corrected using the default source model and 3C286 was then
used to find the phase-calibrator flux densities.  The antenna gains
were calibrated using the phase-calibrator data. Some bad points were
flagged and finally the images were created with natural
weighting. The 1$\sigma_{\mathrm{rms}}$ noise level in the maps was typically
$\sim$50\,$\mu$Jy\,beam$^{-1}$ and the beam was
0\farcs35$\times$0$\farcs$25.

\section{Results}
\subsection{Maser emission}
We successfully mapped a total of 31 out of 33 methanol masers observed
with the EVN. 
We were unable to image G31.156$+$00.045 and G37.479$-$00.105 because of 
the weakness ($<$30\,mJy) of the emission and to strong spike
artefacts in the channels at the maser velocity, respectively.  We
were unable to improve on the MERLIN astrometry for G33.641$-$00.228 
and G35.793$-$00.175 due to a problem with the EVN phase-referencing that 
appeared during the first observing run. 
We created fringe rate maps of the brightest channels of the targets
but still failed to determine the absolute position of these two sources.
The target sources were near zero declination (from $+$0\fdg5 to $+$2\fdg5). 
Furthermore, because of the use of only four EVN telescopes,  
the {\it uv}$-$plane coverage was poor for N$-$S baselines.
It is probable that these factors together with a too long 
phase-referencing cycle time precluded a proper phase calibration. The
position of the third source, G36.115$+$00.552, observed in the first run, 
was easily ascertained during the 4c run when eight antennas were working and the
cycle time between the maser and phase-calibrator was shorter.

The results are summarized in Table \ref{table:5}. The names of the maser
sources correspond to the Galactic coordinates of the brightest spot of each target. 
The absolute coordinates, the LSR velocity (V$_{\rm p}$), and the intensity 
 (S$_{\rm p}$) are given for the brightest spot of each target.
We also indicate the velocity range of emission $\Delta$V. The area
containing all maser emission from each source was parameterised by
measuring the extent of the maser emission along the line given by a
least squares fit to the maser spot distribution (major axis and
position angle) and in the
perpendicular direction (minor axis).  The morphological
class based on the relative positions of methanol maser spots and the
angular separation $\Delta_{\rm MIR}$ of the brightest spot of each
source from the nearest 4.5\,$\mu$m source (see Sect.~4.2) are also
given in Table~\ref{table:5}.

   \begin{figure*}
   \centering
   \includegraphics{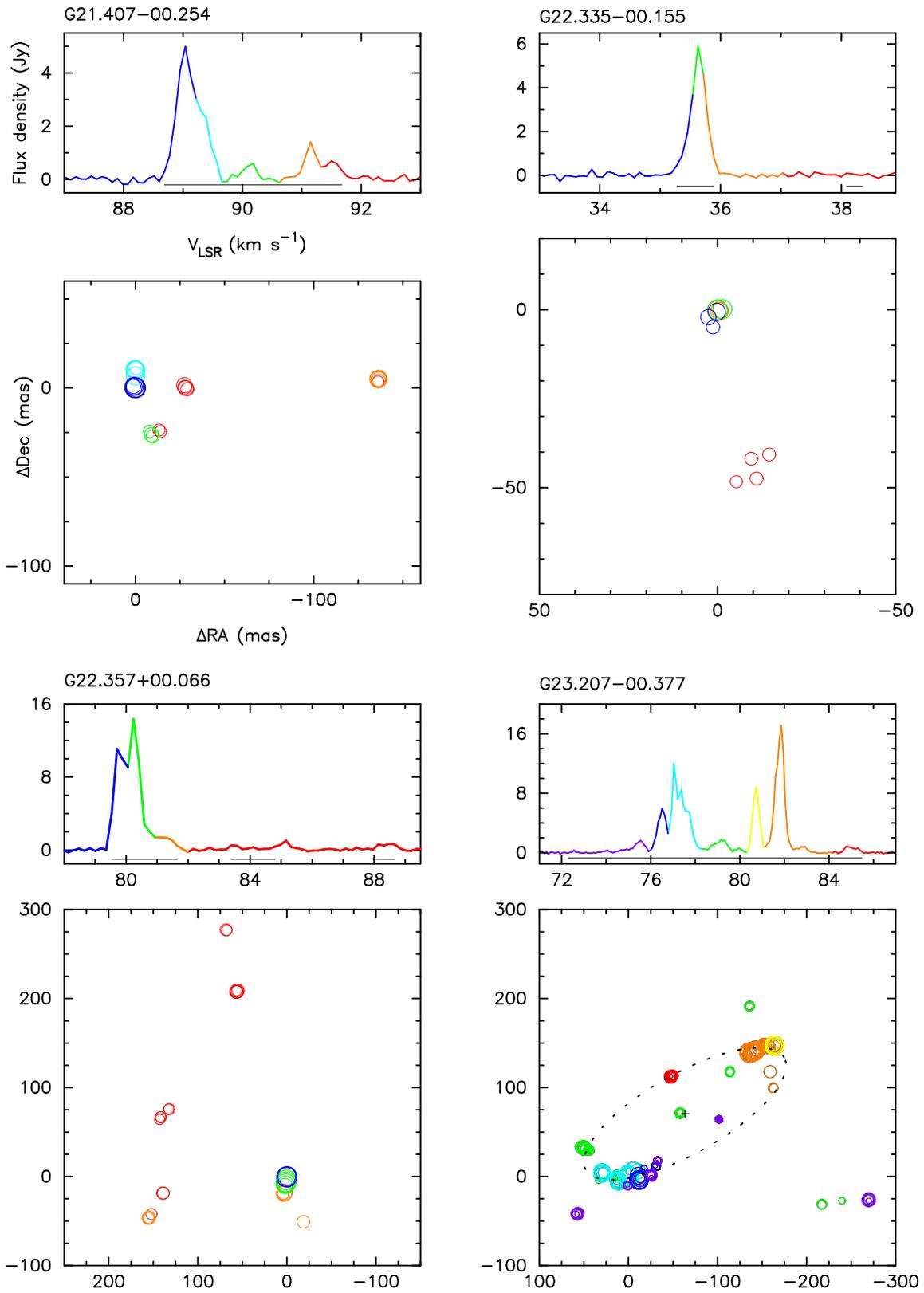}
    \caption{Spectra and maps of 6.7\,GHz methanol maser emission of
          sources detected with the EVN. The names are the Galactic
          coordinates of the brightest spot as listed in Table
          \ref{table:5}. The thin bars under the spectra show the
          velocity ranges of spots displayed. The coordinates are 
          relative to the brightest spots (Table \ref{table:5}). 
          The sizes of circles are proportional to the logarithm 
          of the intensities of maser spots. The colors of circles 
          relate to the LSR velocities as indicated in the spectra, 
          respectively. For the sources with
          ring-like morphologies, the best-fit ellipse and its
          center are marked by a dotted curve and a cross,
          respectively. The crosses coincide (within the
          uncertainties) with \emph{Spitzer} IRAC MIR emission
          (Sect.~4.2).}
\addtocounter{figure}{-1}
         \label{fig1}
   \end{figure*}
   \begin{figure*}
   \centering
   \includegraphics{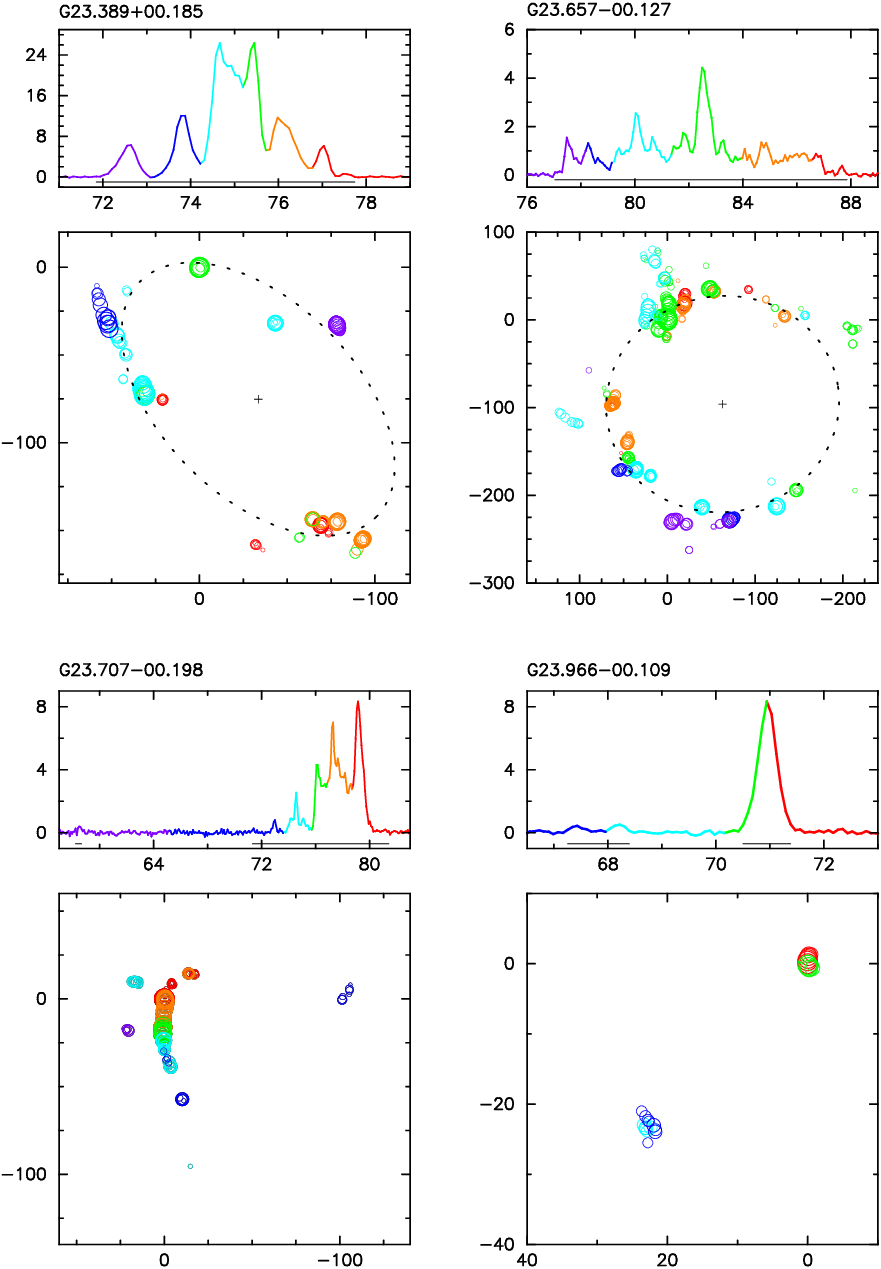}
   \caption{continued.}
\addtocounter{figure}{-1}
   \label{fig2}
   \end{figure*}
   \begin{figure*}
   \centering
   \includegraphics{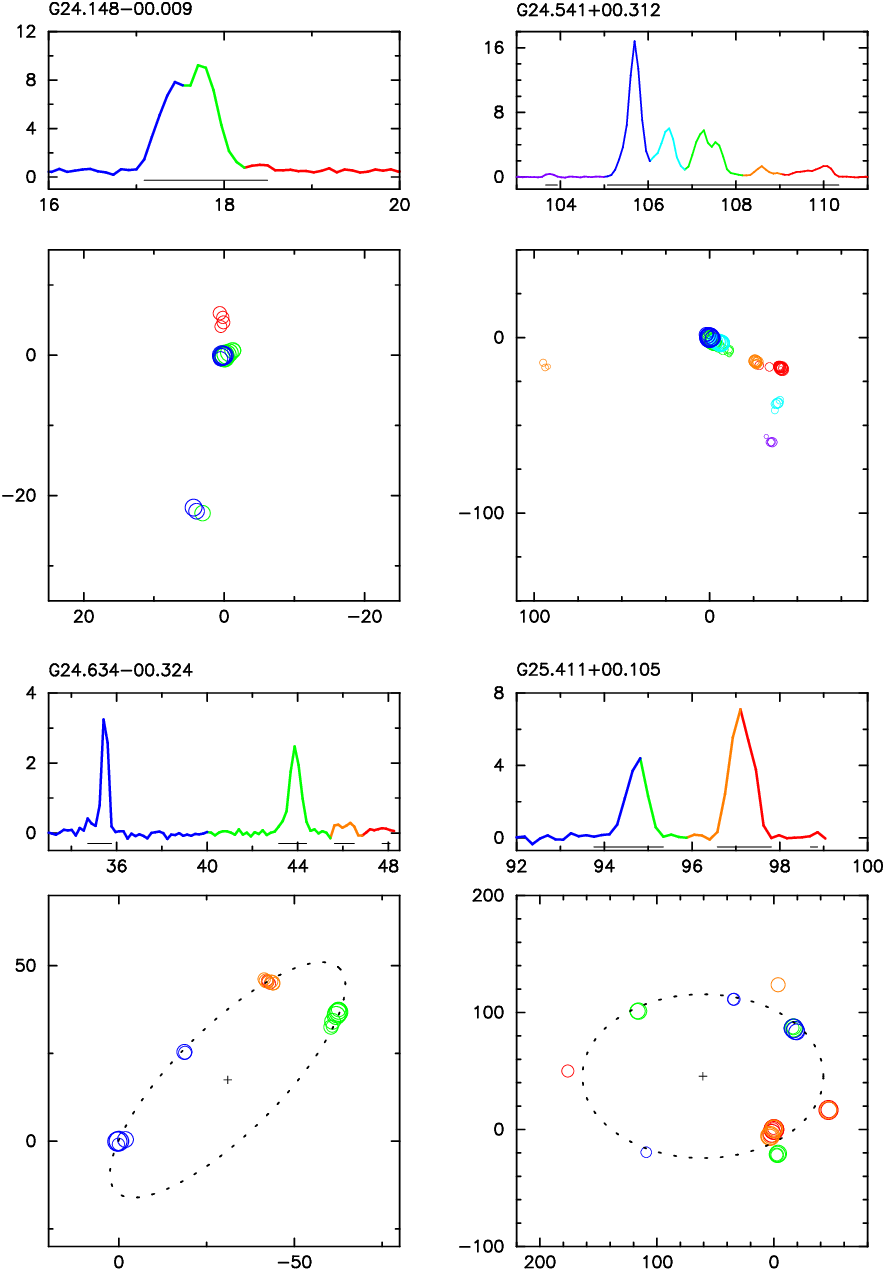}
   \caption{continued.}
\addtocounter{figure}{-1}
   \label{fig3}
   \end{figure*}
   \begin{figure*}
   \centering
   \includegraphics{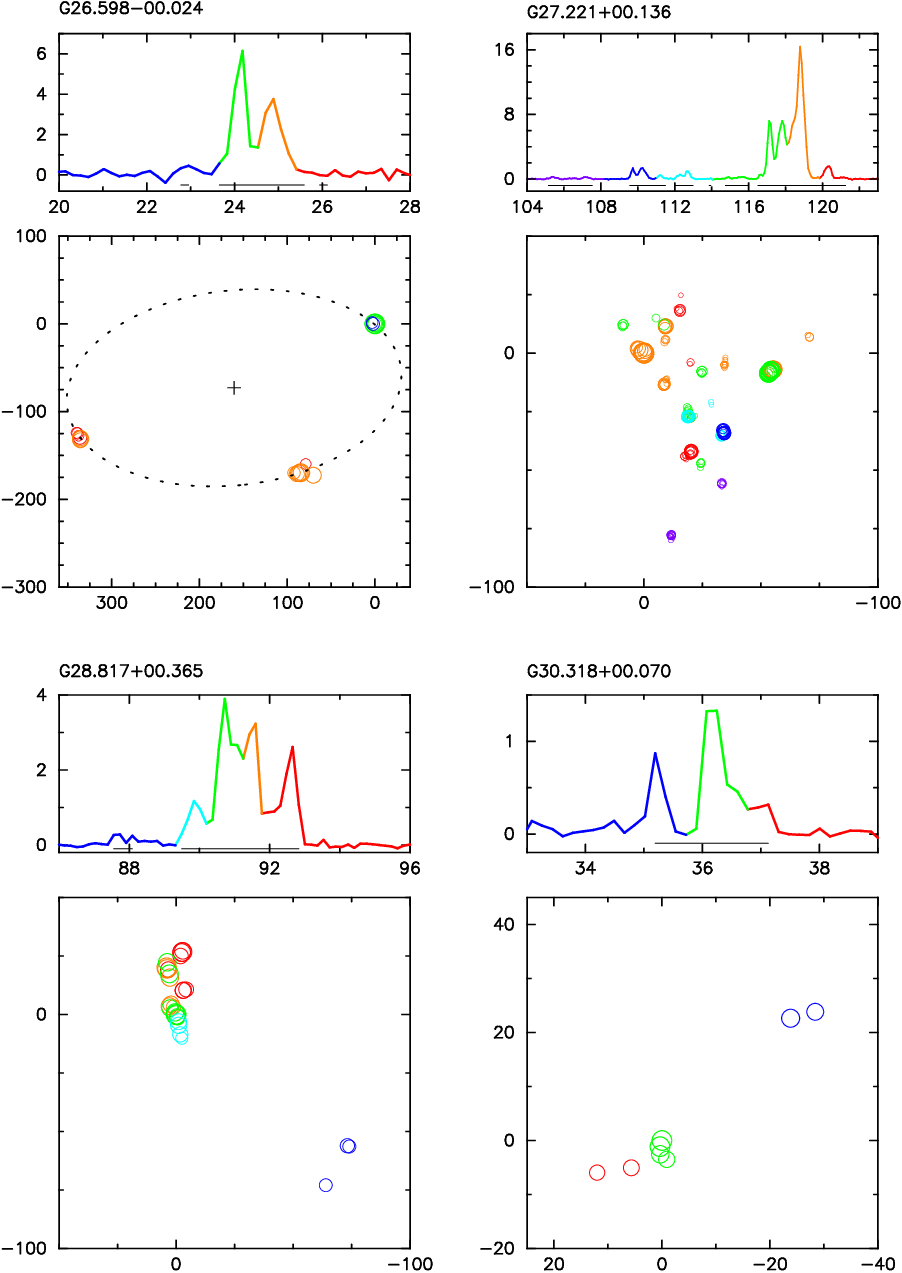}
   \caption{continued.}
\addtocounter{figure}{-1}
   \label{fig4}
   \end{figure*}

   \begin{figure*}
   \centering
   \includegraphics{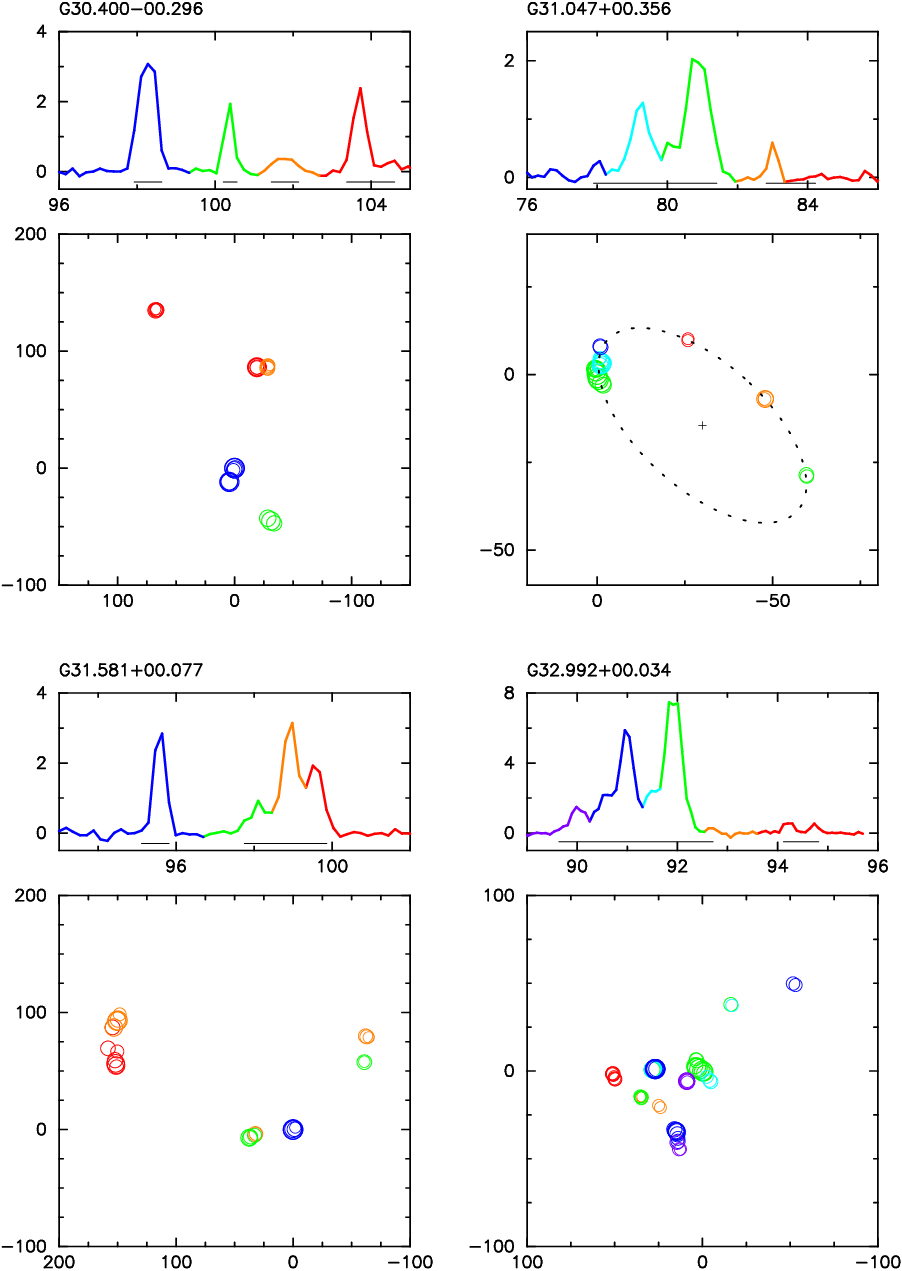}
   \caption{continued.}
\addtocounter{figure}{-1}
   \label{fig5}
   \end{figure*}

   \begin{figure*}
   \centering
   \includegraphics{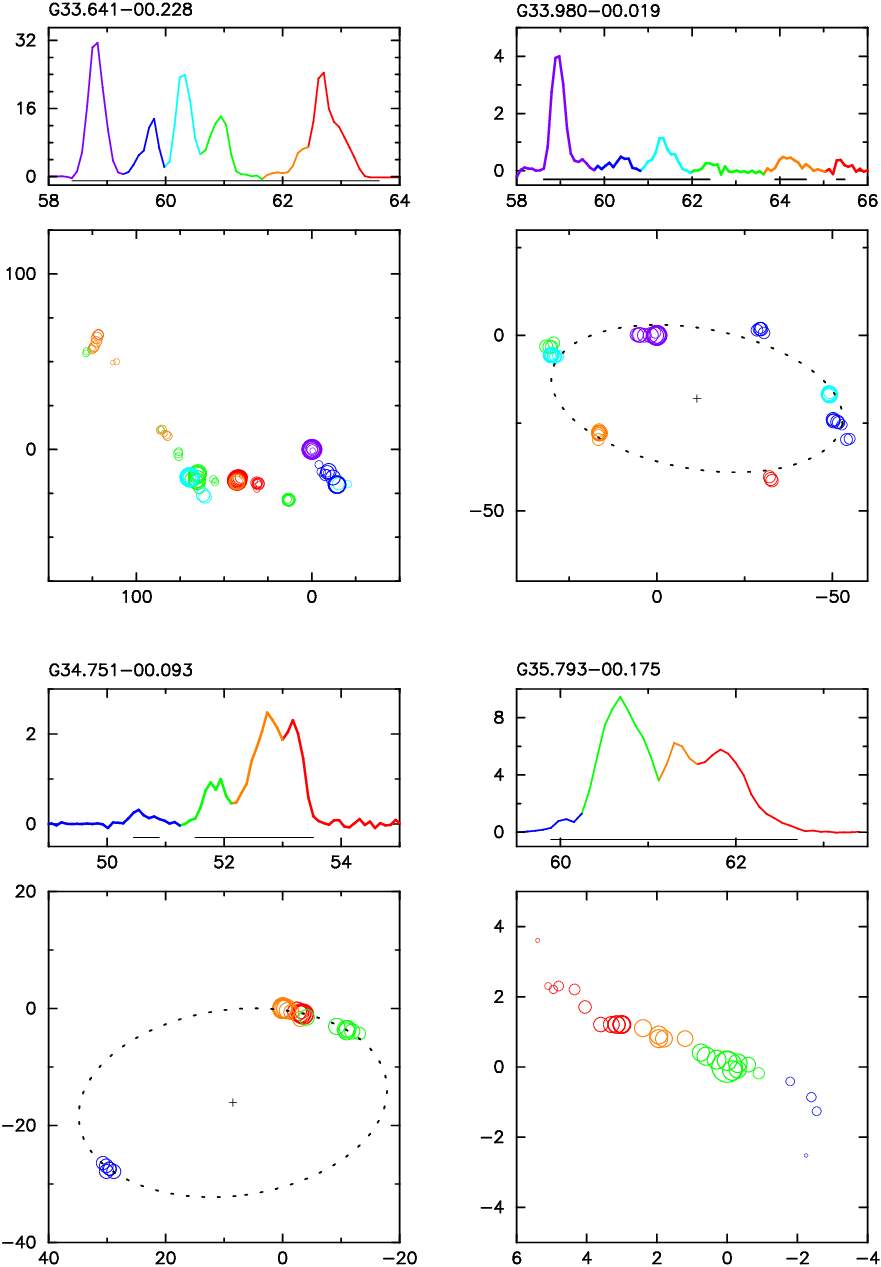}
   \caption{continued.}
\addtocounter{figure}{-1}
   \label{fig6}
   \end{figure*}

   \begin{figure*}
   \centering
   \includegraphics{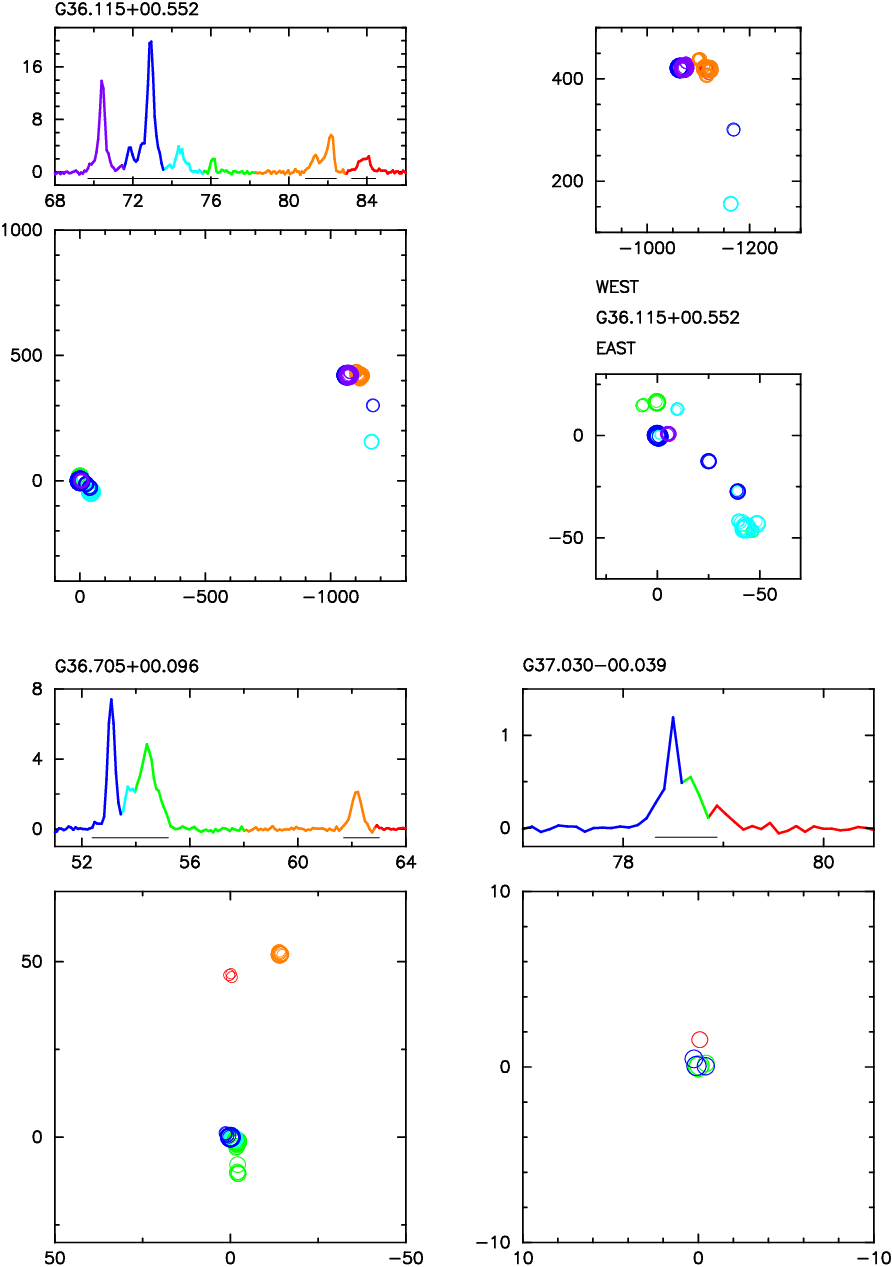}
   \caption{continued.}
\addtocounter{figure}{-1}
   \label{fig7}
   \end{figure*}

   \begin{figure*}
   \centering
   \includegraphics{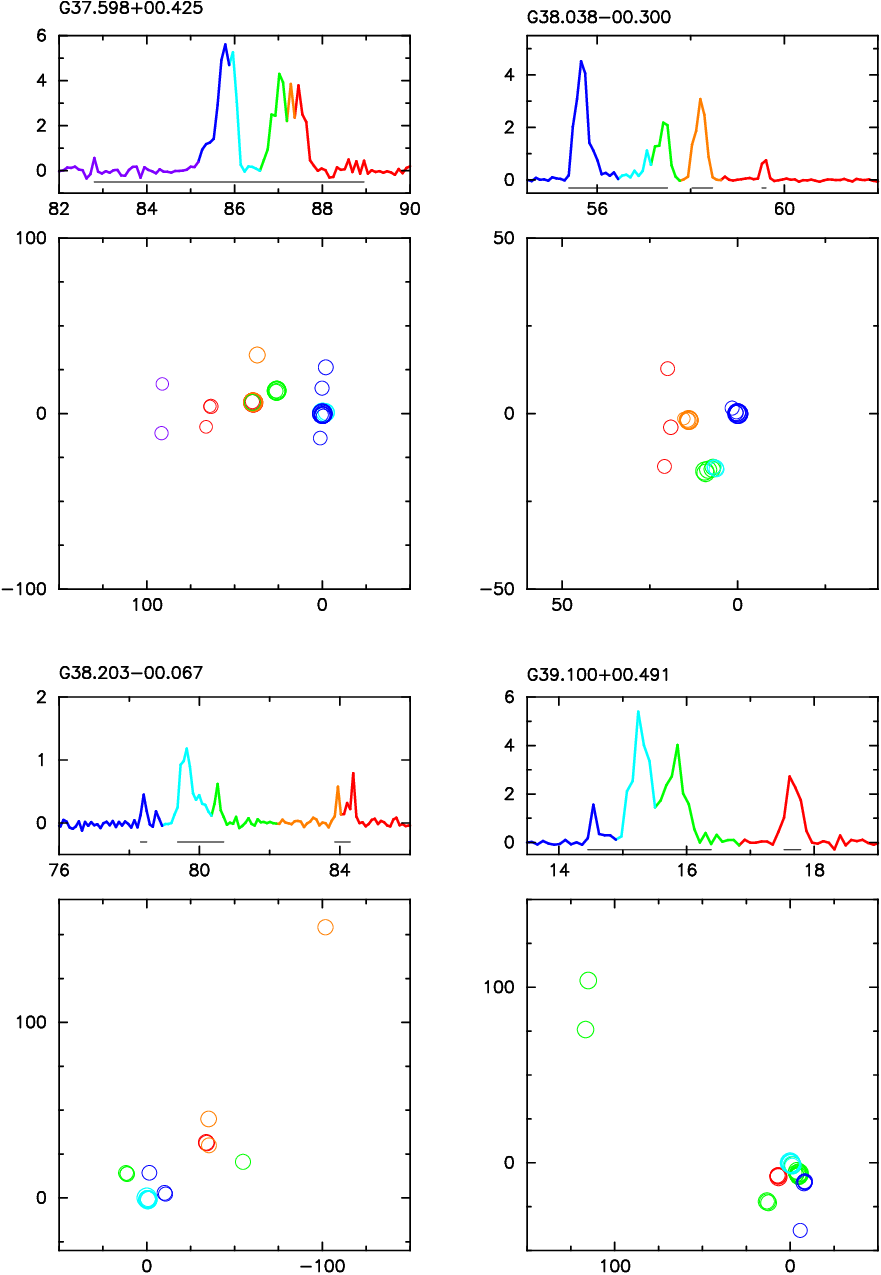}
   \caption{continued.}
   \label{fig8}
   \end{figure*}

In Fig. \ref{fig1}, we present the spectra and distribution
of the methanol maser emission for the 31 imaged targets. The spectra
were extracted from the map datacubes using the AIPS task ISPEC. They
represent the total amount of emission seen in the maps. In order to display
the detailed structures of masers, we show all the spots detected in each of
the individual channel maps. If spots appear at the same positions within 
half of the beamwidth in at least three or two consecutive channels, for
observations with a spectral resolution of 0.09\,km\,s$^{-1}$ and
0.18\,km\,s$^{-1}$, respectively, we refer to them as a {\it cluster}.  The
relevant parameters of all maser clusters for each source are
listed in Table \ref{table:6}: the position ($\Delta$RA, $\Delta$Dec)
relative to the brightest spot (given in Table \ref{table:5}), the
peak  intensity (S$_{\rm p}$), and the LSR velocity (V$_{\rm LSR}$)
of the brightest spot within a cluster. The velocity full-width at 
half-maximum (FWHM) and the fitted peak amplitude (S$_{\rm amp}$) are
given if the spectrum of the cluster has a Gaussian profile.

\begin{table*}
\caption{Results of EVN observations}             
\label{table:5}      
\centering          
\begin{tabular}{lllcccccccc}     
\hline\hline       
\\
Source& \multicolumn{2}{c}{Position (J2000)} & V$_{\rm p}$ & $\Delta$V & S$_{\rm p}$& \multicolumn{2}{c}{Area}& Class$^{**}$& $\Delta_{\rm MIR}$\\
Gll.lll$\pm$bb.bbb& RA(h m s) & Dec(\degr \,\arcmin \, \arcsec)   &
(km\,s$^{-1}$)&(km\,s$^{-1}$) & (Jy\,beam$^{-1}$)&
(mas$\times$mas)&PA(\degr)& & (\arcsec)\\
\hline                    
\\
 G21.407$-$00.254 & 18 31 06.33794 & $-$10 21 37.4108 & 89.0 & 3.00 & 2.76 & 138$\times$39 & $-$87 & C& 0.23 \\
 G22.335$-$00.155 & 18 32 29.40704 & $-$09 29 29.6840 & 35.6 & 3.10 & 1.71 &  49$\times$11 & $+$16 & L& 0.67 \\
 G22.357$+$00.066 & 18 31 44.12055 & $-$09 22 12.3129 & 79.7 & 9.20 &10.54 & 330$\times$174 & $-$5 & C& 0.51 \\
 G23.207$-$00.377 & 18 34 55.21212 & $-$08 49 14.8926 & 77.1 & 13.20 & 9.30 &  313$\times$255&$-$69& R& 0.56 \\
 G23.389$+$00.185 & 18 33 14.32477 & $-$08 23 57.4723 & 75.4 & 6.00 &21.55 &  205$\times$134& $+$59& R& 0.16 \\
 G23.657$-$00.127 & 18 34 51.56482 & $-$08 18 21.3045 & 82.6 & 10.80 & 3.62 &  351$\times$345& $-$82& R&0.50 \\
 G23.707$-$00.198 & 18 35 12.36600 & $-$08 17 39.3577 & 79.2 & 23.30 & 6.06 &  130$\times$110& $-$83& A&0.74 \\
 G23.966$-$00.109 & 18 35 22.21469 & $-$08 01 22.4698 & 70.9 &  4.20 & 5.47 &  35$\times$4   & $-$45& L&0.19 \\
 G24.148$-$00.009 & 18 35 20.94266 & $-$07 48 55.6745 & 17.8 &  1.40 & 3.60  &  28$\times$3   & $-$11&L&0.16 \\
 G24.541$+$00.312 & 18 34 55.72152 & $-$07 19 06.6504 &105.7 &  6.80 & 7.75 &  137$\times$53 & $+$78& A&0.45 \\
 G24.634$-$00.324 & 18 37 22.71271 & $-$07 31 42.1439 & 35.4 & 13.40 & 3.03 &  73$\times$21  & $-$60& R&1.01 \\
 G25.411$+$00.105 & 18 37 16.92106 & $-$06 38 30.5017 & 97.3 &  5.20 & 3.43 &  225$\times$162& $+$79& R&0.64 \\
 G26.598$-$00.024 & 18 39 55.92567 & $-$05 38 44.6424 & 24.2 &  3.30 & 3.04 &  361$\times$152& $-$76& R&0.39 \\
 G27.221$+$00.136 & 18 40 30.54608 & $-$05 01 05.3947 &118.8 & 16.10 &12.54 &  104$\times$79 & $+$6 & C&0.89 \\
 G28.817$+$00.365 & 18 42 37.34797 & $-$03 29 40.9216 & 90.7 &  5.20 & 3.14 &  115$\times$28 & $+$45& A/R&4.71 \\
 G30.318$+$00.070 & 18 46 25.02621 & $-$02 17 40.7539 & 36.1 &  1.90 & 0.52 &  50$\times$6   & $-$50& L&0.87 \\
 G30.400$-$00.296 & 18 47 52.29976 & $-$02 23 16.0539 & 98.5 &  6.70 & 2.77 &  199$\times$97 & $+$47& C/R&2.28 \\
 G31.047$+$00.356 & 18 46 43.85506 & $-$01 30 54.1551 & 80.7 &  6.30 & 1.99 &  68$\times$27  & $+$72& R&1.73 \\
 G31.156$+$00.045$^*$ & 18 48 02.347 & $-$01 33 35.095 &     &       &      &                &      &  &6.62 \\
 G31.581$+$00.077 & 18 48 41.94108 & $-$01 10 02.5281 & 95.6 &  4.80 & 2.72 &  217$\times$105& $+$79& A/R&4.23 \\
 G32.992$+$00.034 & 18 51 25.58288 & $+$00 04 08.3330 & 91.8 &  5.20 & 6.21 &  115$\times$68 & $-$80& C&1.48 \\
 G33.641$-$00.228$^*$ & 18 53 32.563 &$+$00 31 39.180 & 58.8 &  5.30 & 28.3 &  167$\times$61 & $+$66& A&1.22\\
 G33.980$-$00.019 & 18 53 25.01833 & $+$00 55 25.9760 & 58.9 &  6.90 & 3.78 &  89$\times$43  & $+$82& R&0.95 \\
 G34.751$-$00.093 & 18 55 05.22296 & $+$01 34 36.2612 & 52.7 &  3.10 & 1.95 &  49$\times$11  & $-$56& R&0.47 \\
 G35.793$-$00.175$^*$ & 18 57 16.894 &$+$02 27 57.910 & 60.7 &  2.80 & 9.70 &  10$\times$2   & $+$65& L&1.12 \\
 G36.115$+$00.552 & 18 55 16.79345 & $+$03 05 05.4140 & 73.0 & 14.80 &11.74 &1201$\times$297 & $-$79& P&2.42 \\
 G36.705$+$00.096 & 18 57 59.12288 & $+$03 24 06.1124 & 53.1 & 10.60 & 7.58 &  64$\times$18  & $-$16& C&0.32 \\
 G37.030$-$00.039 & 18 59 03.64233 & $+$03 37 45.0861 & 78.6 &  0.70 & 0.69 &  2$\times$1    & $-$15& S&0.61 \\
 G37.479$-$00.105$^*$ & 19 00 07.145 & $+$03 59 53.350 &     &       &      &                &      &  &1.73 \\
 G37.598$+$00.425 & 18 58 26.79772 & $+$04 20 45.4570 & 85.8 &  4.50 & 3.91 &  94$\times$28  & $+$87& C&1.24 \\
 G38.038$-$00.300 & 19 01 50.46947 & $+$04 24 18.9559 & 55.7 &  4.20 & 2.17 &  31$\times$23  & $+$10& C&0.28 \\
 G38.203$-$00.067 & 19 01 18.73235 & $+$04 39 34.2938 & 79.6 &  6.00 & 0.83 &  182$\times$58 & $-$44& C&1.74 \\
 G39.100$+$00.491 & 19 00 58.04036 & $+$05 42 43.9214 & 15.3 &  3.30 & 2.07 &  183$\times$37 & $+$52& C&0.82 \\
\\
\hline\hline                                   
\multicolumn{10}{l}{$^*$ coordinates derived from the single MERLIN baseline data}\\                            
\multicolumn{10}{l}{$^{**}$ class of morphology as described in Sect.~3.3:
S -- simple, L -- linear, R -- ring, C -- complex, A - arched, P -- pair.}\\
\end{tabular}                                                                                
\end{table*}         

\addtocounter{table}{1}

\subsection{Radio continuum emission}
We detected 8.4\,GHz continuum emission in eight of the fields
centered on methanol masers. Table \ref{table:7} lists the continuum
source names (derived from the Galactic coordinates of the 8.4-GHz
peak fluxes), the peak and the integrated intensities, and the angular
size of the radio continuum emission at the 3$\sigma_{\mathrm{rms}}$
level. We also provide the name of the nearest maser from the sample and
the angular separation between the continuum peak and the brightest spot of 
the nearest methanol maser.

\begin{figure*}
   \centering
   \vspace*{0.5cm}
   \includegraphics[scale=0.9]{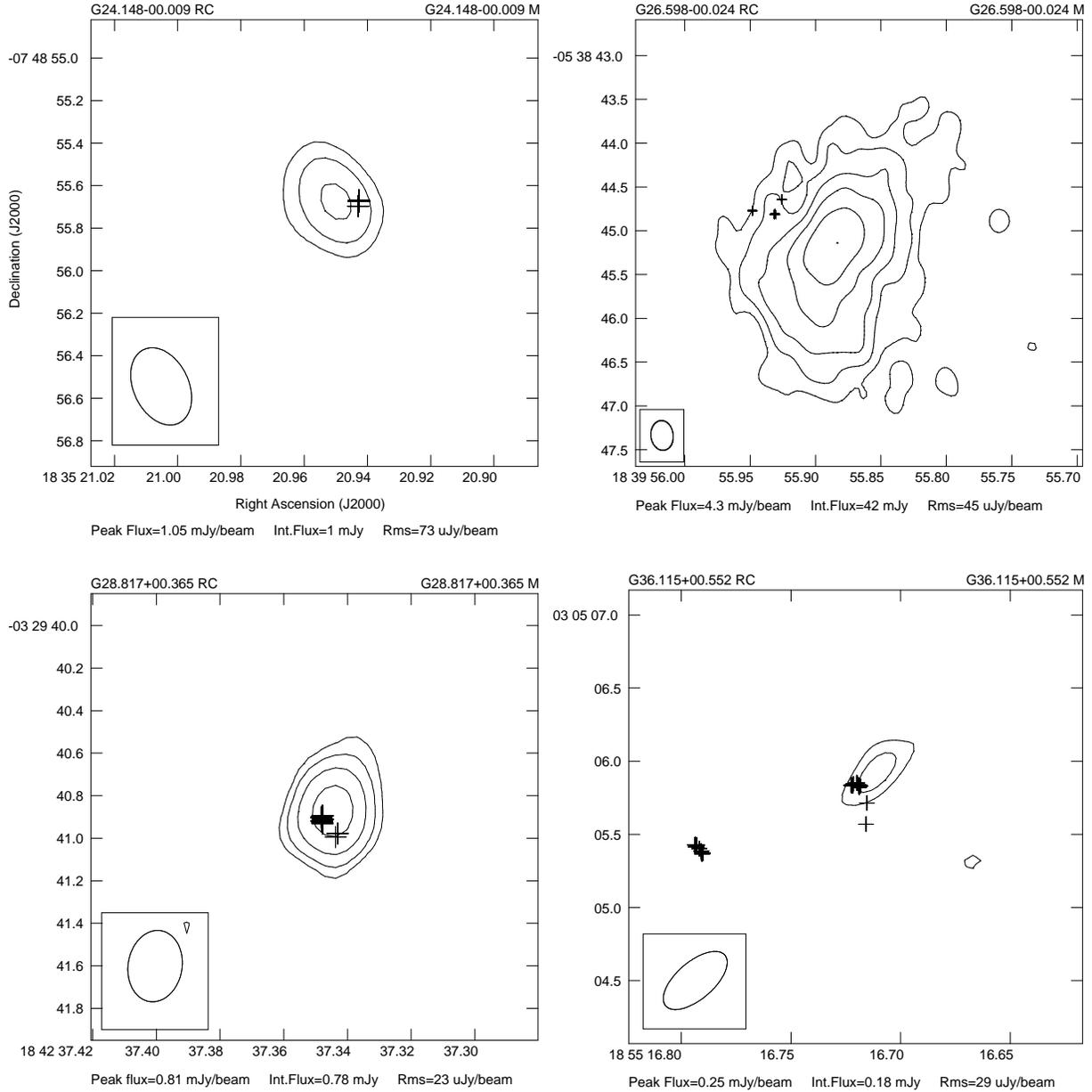}
      \caption{
       The 8.4\,GHz continuum emission towards four methanol maser sites taken
       using VLA on 2007 August 18. The names of radio continuum sources (RC)
       relate to the
       Galactic coordinates of their peak fluxes. The peak and integrated fluxes
       as well as the levels of rms
       (1$\sigma_{\mathrm{rms}}$) are given under each map and the beamsizes are presented at
       the left bottom corners. Contours trace the levels of 3$\sigma_{\mathrm{rms}}$$\times$ 
       (1, 2, 4, 8, 16, 32, 64, ...). 
       Crosses represent the 6.7\,GHz methanol maser spots registered using EVN.
              }
        \label{fig9}
   \end{figure*}

The contour maps of all detections are shown in Figs. \ref{fig9} and
\ref{fig9a}. The majority of sources are single peaked and their
angular sizes range from 0\farcs6 to 3\farcs8. Both G24.148$-$00.009 and
G36.115$+$00.552 have integrated flux densities that are equivalent to their peak
flux densities within the noise, suggesting that these sources are unresolved. The
values given for these sources in Table \ref{table:7} correspond to
the angular size upper limits.  G31.582$+$00.075 is one of the weakest
sources (S$_{\rm p}$=0.43\,mJy\,beam$^{-1}$) but has an
exceptionally complex structure. It is extended
(4\arcsec$\times$3\arcsec) and contains multiple emission peaks.  The
typical upper limit (3$\sigma_{\mathrm{rms}}$) for the fluxes of the 
remaining 22, non-detected sources is 0.15\,mJy\,beam$^{-1}$.

The 6.7-GHz methanol maser emission is found to be within 0\farcs2 of the
8.4-GHz continuum position peaks of G24.148$-$00.009, G28.817$+$00.365, and
G36.115$+$00.552. The maser spots of source G26.598$-$00.024 are 
0\farcs8 from the NE edge of the radio continuum source. Therefore, these
four sources are closely associated with the methanol masers
(Fig. \ref{fig9}).  The continuum object G31.160$+$00.045 is located 
11\farcs9 from the nominal position of the maser source
G31.156$+$00.045, but this maser has a position uncertainty of
10\arcsec\, because we were unable to image it with the EVN (Sect.
3.1), so it may also be associated with the radio continuum.  On the
other hand, the continuum source G31.582$+$00.075 has a separation of
$\sim$9\arcsec\, from the maser G31.581$+$00.077, but the latter has a
position accuracy of a few mas implying that the source and maser 
are unlikely to be associated.
 
We conclude that only 4 (possibly 5) of 30 masers are associated
with radio continuum at 8.4\,GHz. This is consistent with previous
findings (Phillips et al. \cite{phillips98}; Walsh et
al. \cite{walsh98}; Beuther et al. \cite{beuther02}) that the 6.7\,GHz
methanol masers are rarely associated with centimeter wavelength
continuum emission. However, 24-GHz ATCA observations detected 
continuum emission associated with methanol masers toward which no 
continuum at 8.4\,GHz had been previously detected (Longmore et al.
\cite{longmore07}). This opens the possibility of methanol masers 
being associated with hyper-compact H\,{\small II} regions (HC
H\,{\small II}), which are ptically thick at frequencies $<$10\,GHz.

\begin{table*}
\caption{Results of VLA observations at 8.4\,GHz} 
\label{table:7}      
\centering          
\begin{tabular}{lccccccc}     
\hline\hline       
\\
Radio continuum &  S$_{\rm p}$ &S$_{\rm int}$ & \multicolumn{3}{c}{Size}  &
Nearest maser & Separation\\
source &              &             &      Major axis  & Minor axis & PA&&\\
       & (mJy\,beam$^{-1}$)& (mJy)  & (\arcsec) & (\arcsec) & (\degr) && (\arcsec)\\
\hline
\\
G21.385$-$00.254 & 13.00& 65    &3.8 & 1.8 & $+$50         &G21.407$-$00.254& 76.8\\
G24.148$-$00.009 & 1.05 & 1     &0.6$^*$ & 0.4$^*$ & $+$35 &G24.148$-$00.009& 0.11\\
G26.598$-$00.024 & 4.30 & 42    &3.8 & 2.5 & $-$35         &G26.598$-$00.024& 0.80\\
G28.817$+$00.365 & 0.81 & 0.8   &0.6 & 0.5 & $-$20         &G28.817$+$00.365& 0.08\\
G30.330$+$00.090 & 8.70 & 13    &1.3 & 0.8 & $-$25         &G30.318$+$00.070& 85.6\\
G31.160$+$00.045 & 2.40 & 22    &2.0 & 1.5 & $+$70         &G31.156$+$00.045& 11.9\\
G31.582$+$00.075 & 0.43 & 15    &4.0 & 3.0 & 0             &G31.581$+$00.077& 9.00\\
G36.115$+$00.552 & 0.25 & 0.2   &0.7$^*$ & 0.3$^*$ & $-$45 &G36.115$+$00.552& 0.20\\
\\
\hline\hline                                   
\multicolumn{8}{l}{$^*$ angular size upper limits (see Sect.~3.2)}\\
\end{tabular}                                                                                
\end{table*}

\onlfig{3}{
   \begin{figure*}
   \centering
   \vspace*{0.5cm}
   \includegraphics[scale=0.9]{continuum_online_3rms.epsi}
      \caption{The 8.4\,GHz continuum emission detected using VLA on 2007 August 18.
      The names of radio continuum sources (RC) are the Galactic coordinates of 
      peak fluxes. The peak and integrated fluxes as well as the levels of rms
       (1$\sigma_{\mathrm{rms}}$) are given under each map and the beamsizes
       are presented at the left down corners. Contours trace the levels of
       3$\sigma_{\mathrm{rms}}$$\times$(1, 2, 4, 8, 16, 32, 64, ...).}
         \label{fig9a}
   \end{figure*}
}
\subsection{General properties of the 6.7\,GHz methanol masers}
In 31 sources, we detected a total of 1934 maser spots that form 333 clusters.
The spectral profiles of 265 (80\%) clusters are well fitted with
a Gaussian. The mean FWHM is 0.41$\pm$0.01\,km\,s$^{-1}$ and
the median value is 0.37\,km\,s$^{-1}$. This is consistent with results from
single dish spectra at\, $\sim$0.05\,km\,s$^{-1}$ resolution 
(Menten \cite{menten91}; Caswell et al. \cite{caswell95}). 
Nineteen sources have complex spectra, that is indicative of spectral blending, so that
the line width of individual features cannot be properly determined solely
from the spectrum.

 We compared the basic parameters of the spectra and distributions of
all masers from the sample. However, 
we did not find any relationships between the line parameters such as
FWHM, brightness temperature, velocity range of the maser emission, and
the size and geometry of the maser region.

The sources show a wide diversity of structures. The following types
of morphology can be identified (Table \ref{table:5}):

{\it Simple} -- the emission appears in a narrow velocity range 
($\Delta$V=0.7\,km\,s$^{-1}$) as a single peaked spectrum. The maser spots
form one cluster of size smaller than a few mas. G37.030$-$00.039 is the
only source with these properties. Its spectrum is obviously blended.

{\it Linear} -- the maser spots form a line in the plane of the
sky. The angular extent of these maser structures ranges from 9 to
54\,mas. In some sources (G30.318$+$00.070, G35.793$-$00.175) a
monotonic velocity gradient is clearly seen. There are five linear
sources in the sample.

{\it Ring} -- this morphology appears to be ubiquitous in our
sample. The distributions of no less than nine sources display a ring
structure. Using the GNU Octave script developed by Fitzgibbon et
al. (\cite{fitzgibbon99}), we fitted an ellipse to the spatial
positions of the maser spots for each source. The results are
summarized in Table \ref{table:8}. The semi-major (a) and semi-minor
(b) axes range from 27 to 192\,mas and from 15 to 128\,mas,
respectively. The average size of the semi-major axis and the standard
dispersion in the mean is 89$\pm$20\,mas. The eccentricity (e) of the
best-fit model ellipses ranges from 0.38 to 0.94. The average
eccentricity and the standard dispersion of the mean is
0.79$\pm$0.06. The emission spans a modest velocity range of
(3.1$-$13.4\,km\,s$^{-1}$). All nine sources possess MIR counterparts
that coincide with the ellipse center to within less than
2\farcs5 (Table \ref{table:5}). In these objects it is very likely 
that ring-like maser emission surrounds a central embedded star
(see Sect.~4.2). Three other sources (G28.817$+$00.365, G30.400$-$00.296,
G31.581$+$00.077) have a ring-like morphology,
although the separation between the MIR candidate counterpart and the ellipse
center is greater than 2\farcs5 (Table \ref{table:5}). 
This is probably caused by the larger uncertainties in the maser
positions, since all three sources are at declinations near 0\degr.
These are assigned a tentative classification of the 
ring-like class in Table \ref{table:5}.

{\it Arched} -- maser spots are distributed along an arc of between 70 and 
220\,mas in length. The entire structure may show a systematic  velocity gradient. 
Three (or possibly five) sources exhibit this morphology. 

{\it Complex} -- 9 (possibly 10) sources, i.e. about one third of the sample, 
do not show any regularities in their spatial and velocity distributions.
These sources vary greatly in size from 31$\times$23\,mas$^2$ to 
330$\times$174\,mas$^2$.

{\it Pair} -- this class was defined by Phillips et
al. (\cite{phillips98}), comprising two maser groups separated by
$\sim$1\,arcsec with $\ge$10\,km\,s$^{-1}$ difference in velocity. The
major axes of individual clusters are perpendicular to the line joining
them. In our sample, we found only one source with such a morphology.

The most striking aspect of this study is that we find that the
commonest morphology of sources with a systematic maser structure is a
ring-like distribution of emission, seen in $\sim$29\% of
objects. These rings probably surround young massive objects (see
Sect.~4). A similar proportion ($\sim$29-32\%) of sources possess a
complex morphology.  Linear sources with a monotonic velocity gradient
are relatively rare in the sample ($\sim$16\%).

\subsection{Individual sources}
\label{sec:individual}
This section presents comments on individual sources, including
additional observational data relevant to the main aims of this
paper. 
If not stated otherwise, we present the linear sizes of masers 
derived using the near (and far) kinematic distances given in Szymczak et 
al. (\cite{szymczak05}).

{\bf G22.357$+$00.066.} The ATCA observations detected three maser
spots (Walsh et al.\,\cite{walsh98}), while the EVN revealed 31 spots
in 10 clusters. The strongest emission detected with both
interferometers, at 80.0\,km\,s$^{-1}$, coincides to within 1\farcs7.  We
detected new emission at close to 88.5\,km\,s$^{-1}$, 0\farcs20 - 0\farcs27
north of the brightest spot, but a 77.0\,km\,s$^{-1}$ spot reported by
Walsh et al. (\cite{walsh98}) was not redetected.
 
{\bf G23.657$-$00.127.} The parallax of this source  was
measured (Bartkiewicz et al. \cite{bartkiewicz08}), showing that the
circular distribution of masers has a linear radius of 405\,AU, which
differs significantly from the sizes inferred from the kinematic
distances.  The source was observed at three epochs (Tables
\ref{table:2}, \ref{table:4}) and the ring-like morphology clearly
persisted over time spans of 1.25 and 2.5\,yrs. A detailed description
of the brightness variability in the source will be presented in a
forthcoming paper.

{\bf G23.707$-$00.198.} Walsh et al. (\cite{walsh98}) detected 7
masers in a velocity range of 74.9$-$81.4\,km\,s$^{-1}$, randomly
scattered over a 0\farcs15 area.  The first epoch of EVN maps of this
source (run 2) detected 23 clusters (140 spots) of which 19 form a
71\,mas (corresponding to 360/750\,AU for near/far kinematic
distance, respectively) long arc in the NS direction, which has a
velocity span of 8\,km\,s$^{-1}$. A clear velocity gradient is seen in
the overall arched distribution. The remaining clusters (all
blue-shifted) are located $\sim$100\,mas to the west (two clusters)
and $\sim$20\,mas to the east (two clusters), relative to the
brightest spot. All four clusters are weak and were not
detected at a later epoch (run 3a),  but this data had poorer 
sensitivity. The brightest methanol maser component (Table
\ref{table:5}) coincides in position (within 82\,mas) and velocity
(within 0.1\,km\,s$^{-1}$) with weak (60\,mJy\,beam$^{-1}$) H$_2$CO maser
emission at 4.8\,GHz (Araya et al.\,\cite{araya06}). We note that this is
well within the absolute positional accuracy of the H$_2$CO
maser. Both masers lie at the edges of two probable H\,{\small II}
regions (Araya et al.\,\cite{araya06}, their Fig. 2) with a peak
intensity of 6.1\,mJy\,beam$^{-1}$ at 5\,GHz (VLA
C-configuration). Our VLA A-configuration data at 8.4\,GHz do not show
any emission above the 0.15\,mJy\,beam$^{-1}$ sensitivity limit, nor
was this source detected at 8.6\,GHz with a 2\,mJy\,beam$^{-1}$ limit
(Walsh et al.  (\cite{walsh98}). Therefore, the H\,{\small II} regions
are intrinsically weak at frequencies $>$5\,GHz or they are possibly
resolved at subarcsec angular resolution.

{\bf G25.411$+$00.105.} Beuther et al. (\cite{beuther02}) observed
this source with the ATCA and found only two components at velocities of 
94 and 97\,km\,s$^{-1}$ at positions that coincide to within 0\farcs2
with the brightest spots in the EVN maps.  We detected 30 spots,
probably because of our $\sim$50 times higher sensitivity, although
variability may also be involved. The distance of 9.5\,kpc (Sridharan
et al. \cite{sridharan02}) implies that the linear radius of the ring
distribution is 980\,AU.

{\bf G26.598$-$00.024.} The maser is located 0\farcs85 from a cometary
H\,{\small II} region (Fig. \ref{fig9}) with a spectral index of 0.23
between 1.4 and 5\,GHz (Becker et al.\,\cite{becker94}).  This
corresponds to linear distances of 1530(11400)\,AU.  The flux density
of 4.4\,mJy at 8.4\,GHz, compared with 55\,mJy at 5\,GHz, implies that
the turnover frequency is near 5\,GHz. The methanol maser probably
forms behind a shock front induced by the H\,{\small II} region. The
strongest maser component, at 24.2\,km\,s$^{-1}$, coincides in
velocity with a 24.3\,km\,s$^{-1}$ absorption feature ($-$59.8\,mJy)
in the H$_2$CO line at 4.8\,GHz (Sewilo et al.\,\cite{sewilo04}).

{\bf G36.115$+$00.552.} The brightest component of the NW maser
structure is 0\farcs2 away from the weak point continuum source at PA
= 120\degr \,(Fig. \ref{fig9}).  The shape of this complex suggests the
existence of an outflow. However, the kinematics are not consistent
with an outflow model (Sect.~4.2.3), nor do CO (2-1) line maps at
29\arcsec \,resolution detect any molecular outflow (Zhang et
al.\,\cite{zhang05}).

\begin{table}
\caption{Parameters of ellipses fitted to the maser spot
distributions.}
\label{table:8}      
\centering          
\begin{tabular}{lccccc}     
\hline\hline       
\\
Source & Centre &\multicolumn{2}{c}{Semi--axes} & PA$^{**}$& e\\
       & $\Delta$RA,$\Delta$Dec$^*$   &a  & b& &\\
       & (mas,mas) & (mas) & (mas) & (\degr)& \\
\hline
\\
G23.207$-$00.377 &$-$62,    71& 126 & 45  & $-$60 & 0.93 \\
G23.389$+$00.185 &$-$34, $-$75& 95  & 56 & $+$45 & 0.81 \\
G23.657$-$00.127 &$-$69, $-$93& 133 & 128 & $-$10 & 0.38 \\
G24.634$-$00.324 &$-$31,    17& 45  & 15  & $-$45 & 0.94 \\
G25.411$+$00.105 &   61,    46& 103 & 70 & $+$90 & 0.73 \\
G26.598$-$00.024 &  161, $-$73& 192 & 111 & $-$84 & 0.81 \\
G31.047$+$00.356 &$-$30, $-$15& 37  & 18  & $+$47 & 0.87 \\
G33.980$-$00.019 &$-$11, $-$18& 42  & 20  & $+$80 & 0.88 \\
G34.751$-$00.093 &    9, $-$16& 27  & 16  & $-$83 & 0.80 \\
\\
\hline\hline                       
\multicolumn{6}{l}{$^*$ coordinates relative to the brightest spots 
as listed in Table \ref{table:5}.}\\            
\multicolumn{6}{l}{$^{**}$ the position angle of semi-major axis
(north to east).}\\
\end{tabular}                                                               
\end{table}

\section{Discussion}
\subsection{Kinematic models of the origins of methanol masers}
The diverse morphologies of methanol masers indicate that there is no 
straighforward explanation of the origin of this emission, as 
discussed previously (Norris et al. \cite{norris93}; Phillips et
al. \cite{phillips98}; Minier et al. \cite{minier00}). The main
hypotheses for the origin of methanol masers assume that they
originate in a circumstellar disc or torus, in outflows or a shock colliding
with a rotating molecular cloud. The results that we obtained by applying 
the existing models to our data are summarized below.

\subsubsection{Rotating and expanding ring} 
\label{sec:ring}
Ring-like masers, which are prevalent in the present sample,
may represent an inclined disc or torus around a
massive protostar or young star. There is a tendency for flatter
structures to have a larger velocity width, in all sources apart from
G23.657$-$00.127, (see Table \ref{table:5}), which suggests that we 
observe the effects of inclination and all motion is 
intrinsic to the plane of the ring.  To test this
hypothesis, we applied the model of a rotating and expanding thin disc
(Uscanga et al. \cite{uscanga08}) to the nine ring-like masers. First,
the coordinates of the spots (x$_{\rm j}$, y$_{\rm j}$) were
transformed to a reference system (x$_{\rm j}'$, y$_{\rm j}'$) in
which the origin was the center of the ellipse and the x'--axis was
directed along the major axis of the projected ellipse (see Fig.~1 in
Uscanga et al. \cite{uscanga08}). We then attempted to reproduce the
kinematics using the LSR velocities (V$_{\rm LSR}$) of the maser spots 
to determine rotation (V$_{\rm rot}$), expansion (V$_{\rm
exp}$), and systemic (V$_{\rm sys}$) velocities of each source. The
solutions were based on the minimisation of the $\chi_{\rm V}^2$ function 
expressed in Eq.~(8) by Uscanga et al. (\cite{uscanga08})
\begin{eqnarray}
{\rm \chi_V^2=\frac{1}{N-3}\sum\limits_{j=1}^N \frac{1}{\sigma_V^2}\Big(V_{{\rm
LSR},j}-V_{\rm sys}-\frac{x_j'}{a}V_{\rm rot} \sin i-\frac{y_j'}{a}V_{\rm
exp} \tan i\Big)^2}, \nonumber
\end{eqnarray}
where $\sigma_{\rm V}$ was the spectral resolution of the observations
corresponding to the uncertainty in the LSR velocity. The inclination angle
is the angle between the line-of-sight and the normal to the ring
plane, which is defined to be i$={\rm acos} (\frac{\rm b}{\rm a})$. The
semi--major and semi--minor axes (a,b) were taken from Table
\ref{table:8}.  We note, that we cannot determine the sign of
the inclination angle from the data available and the
direction of the rotation is therefore ambiguous, nor distinguish between
outflow and contraction. Additional information (e.g., 
spectroscopic and interferometric observations of molecular clouds at
mas resolution) are necessary to solve these questions.  For the
purpose of the model, we assumed that the brightest and the most
complex half of the ellipse is closer to the observer. The results of
fitting are summarized in Table \ref{table:10} and an example of the
fit of the model to the data, for G33.980$-$00.019, is presented in
Fig. \ref{fig16}.

We note that in general the expansion/infall velocity is higher than 
the rotation component in the majority of sources, as can be 
 clearly seen in the maser spot distributions
(Fig. \ref{fig1}). If rotation velocity was instead higher than that of
expansion or infall, the extreme values of 
velocities could be produced where the major axis and ellipse
intersect (at a position angle of 0\degr\, from the major axis). In four
of nine sources, the opposite result is found that higher blue- or red-shifted
velocities appear where the minor axes intersect the ellipses (see
plots of G23.207$-$00.377, G23.389$+$00.185, G24.634$-$00.324, and
G25.411$+$00.105). The average position angle of the most extreme
velocity with respect to the major axis in all nine rings, and the
standard dispersion in the mean, is 52\degr$\pm$11\degr.  This suggests
that the masers originate in the zone where the radial motions exist
and expansion/infall plays a role. This could occur at the interface
between the disc/torus and outflow. A similar result was 
reported for the archetypical object Cep~A (Torstensson et
al. \cite{torstensson09}), where the major axis of the elliptical
distribution of 6.7\,GHz methanol masers is perpendicular to the
bipolar outflow. Their LSR velocity distributions show similar
characteristics in that expansion or contraction dominates over the
rotation.

\begin{table}
\caption{Parameters derived by fitting kinematics of the rotating
and expanding disc model. The signs $+$ and $-$ of the rotation and expansion
velocities refer to the clockwise or anti-clockwise rotation and 
outflow or inflow for positive i.  Both rotation and flow 
are reversed in a case of negative i. 
For each source, both signs together with the sign of i could be reversed
since our model does not give the directions unambiguously. 
}
\label{table:10}      
\centering          
\begin{tabular}{lrrccc}     
\hline\hline       
\\
Source & V$_{\rm rot}$ &V$_{\rm exp}$ &V$_{\rm sys}$ & i&$\chi_V^2$ \\
       & (km\,s$^{-1}$)& (km\,s$^{-1}$)&(km\,s$^{-1}$)& (\degr)& \\
\hline
\\
G23.207$-$00.377 & $-$1.17 &$+$3.96& 79.46& $-$69& 168\\
G23.389$+$00.185 & $-$1.26 &$-$1.71& 74.91& $+$54& 101\\
G23.657$-$00.127 & $+$7.29 &$-$2.61& 81.90& $+$16& 641\\
G24.634$-$00.324 & $+$8.64 &$-$2.25& 38.96& $+$71& 544\\
G25.411$+$00.105 & $+$0.09 &$-$1.17& 95.84& $+$47& 206\\
G26.598$-$00.024 & $+$0.81 &$-$0.81& 24.58& $-$55&  30\\
G31.047$+$00.356 & $+$4$\times$10$^{-7}$ & $-$2.43& 80.68& $-$61& 211\\
G33.980$-$00.019 & $-$0.45 &$+$2.97& 61.86& $-$62& 123\\
G34.751$-$00.093 & $-$1.17 &$-$2.88& 51.17& $+$53&  22\\
\\
\hline\hline                                   
\end{tabular}                                                                                
\end{table}

   \begin{figure}
   \centering
   \vspace*{0.5cm}
   \includegraphics{model_g33.980.epsi}
      \caption{Velocity of the maser spots in G33.980$-$00.019 versus
azimuth angle measured from the major axis (north to east). The open
circles represent the data and are proportional to the logarithm of
the flux densities. The sinusoidal line represents the best-fit
kinematical model of a rotating and expanding disc (with infintesimal 
thickness) using the parameters listed in Table \ref{table:10}.}
         \label{fig16}
   \end{figure}

\subsubsection{Linear maser as edge-on disc?}
A thin disc seen edge-on would appear to have a linear morphology.
Norris et al. (\cite{norris98}) argued that maser radiation propagates
most strongly in the plane of the disc due to the greater column
depth, to explain why so many methanol masers with linear
morphology appeared in the data then available.

We do not confirm this selection effect and note that the increased
sensitivity detects more complex structures of masing regions. Only
16\% of our sample are linear masers (compared to 29\% ring-like
masers) and they are also not the brightest, although it is possible
that if there is strong expansion or infall, this would produce 
a steeper velocity gradient in edge-on discs and possibly
make the masers fainter. 

We calculated the mass that each linear maser structure would contain if
it originated in a disc in Keplerian rotation, using a
method similar to that of Minier et al. (\cite{minier00}). Assuming
that the masing area corresponds to the diameter of the disc, the
average central mass of these five linear structures is
0.12\,M$_\odot$ or 0.44\,M$_\odot$ for the near or far kinematic
distances, respectively. These subsolar values are very unlikely
for massive protostars. We agree with Minier et al. (\cite{minier00})
that the underlying assumption is wrong and we do not detect the
full diameter of the disc. However, if we assume that the true extent of the
linear masers is similar to the average size of the major axis  
of the nine ellipses (188\,mas), this implies a mean central mass
of 76\,M$_\odot$ or 190\,M$_\odot$ for the near or far kinematic
distance, respectively. These values seem unrealistically high. Another
solution is that the masers are not bound by Keplerian rotation and 
we argue that the most likely explanation is that
the linear morphology results from a different scenario. Linear
structures with ordered velocity gradients may be produced readily by
geometric effects in outflows. It seems significant that Szymczak et
al. (\cite{szymczak07}) detected molecular line emission from HCO$^+$,
CO, and CS towards these five sources using the IRAM 30\,m telescope,
supporting the outflow scenario. In addition, De Buizer et
al. (\cite{buizer09}) imaged SiO outflows towards five methanol masers
with linear morphologies.  They found that the spatial orientations of
the outflows were inconsistent with the methanol masers tracing
discs. Linear masers produced by outflows seemed to provide a much more
plausible scenario. Finally, we also note that the linear masers have
a smaller extent than most other maser structures (Table
\ref{table:5}). In particular, the entire G35.793$-$00.175 structure
is only 10\,mas long corresponding to the typical size of an
individual maser cluster in other sources. We conclude that most of
the linear masers are more likely to be associated with outflows than
with edge-on discs, although it is possible that more sensitive
observations might indicate that some are part of ring-like or
other structures. G25.411$+$00.105 (see Sect.~\ref{sec:individual})
provides such an example, since Beuther et al. (\cite{beuther02})
found only two maser spots, whilst in the present study we detected
30 spots, forming a ring-like morphology.

\subsubsection{Propagating shock front}
Dodson et al. (\cite{dodson04}) proposed another model for linear
methanol masers.  A low speed planar shock propagating through the
rotating molecular cloud would produce a linear spatial distribution
of maser spots if the shock was propagating close to perpendicular to
the line of sight. The linearity would be distrupted where the shock
interacted with density perturbations in the star-forming regions. The
main diagnostic for this scenario is the perpendicular orientation of
velocity gradients within individual clusters with respect  to the
main large-scale velocity gradient. We analyzed the internal gradients
of clusters and found this behaviour in three out of five masers with
linear morphologies (G22.335$-$00.155, G23.966$-$00.109, and
G30.318$+$00.070). In addition, the arched source, G33.641$-$00.228,
shows similar characteristics in four (out of six) clusters, which have
internal velocity gradients perpendicular to the longest axis of the
overall structure. All these masers have young massive objects in
close proximity less then 1\farcs22 away 
(Sect.~\ref{sec:mir}), which could be responsible for the external shocks.
Proper motions studies are needed to verify this scenario.

\subsubsection{Bipolar outflow}
The bipolar outflow model for H$_2$O masers associated with a
high--mass young stellar object was proposed by Moscadelli et al.
(\cite{moscadelli00}) and confirmed in IRAS20126$+$4104 by proper
motion studies (Moscadelli et al. \cite{moscadelli05}).  We also
tested this model for all sources in this study.  The assumptions of the
model are as follows: masers originate in the surface of a conical bipolar
jet due to the interaction between the ionised jet and the surrounding
neutral medium, and the velocity of a maser spot, V$_{\rm o}$, is directed
radially outward from the central star at a constant value. The
center of the coordinate system is at the vertex of the cone, the
z--axis is along the line of sight, and the x--axis coincides with the
projection of the outflow on the plane of the sky (see Fig.~4 in
Moscadelli et al. (\cite{moscadelli00})). Taking the central velocity
of the maser emission (V$_{\rm c}$) to be the systemic LSR velocity, we
minimized the $\chi^2$ function as expressed in Eq.~(3) of Moscadelli
et al.  (\cite{moscadelli00})
\begin{eqnarray}
{\rm \chi^2= \sum\limits_{j=2}^n \bigg(\frac{V_z^j}{V_z^1}-\frac{V_{LSR}^j-V_c}
{V_{LSR}^1-V_c}\bigg)}^2, \nonumber
\end{eqnarray}
where n is the number of spots, V$_{\rm z}^{\rm j}$ is the velocity of
the spot \#j calculated using Eqs.~(1) and (2) from Moscadelli et
al. (\cite{moscadelli00}), and ${\rm V_{LSR}^j}$ is the observed LSR
velocity of the spot.

\begin{table}
\caption{Parameters derived from fitting the biconical outflow model
of Moscadelli et al. (\cite{moscadelli00}).}
\label{table:11}
\centering
\begin{tabular}{lcccccc}
\hline\hline
Source   & $\Delta$RA,$\Delta$Dec$^*$ & V$_{\rm o}$ & PA & $\Psi$ & $\theta$ &$\chi^2$\\
     & (mas, mas)  & (km\,s$^{-1}$)& ($^o$) & ($^o$) & ($^o$) & \\
\hline
\\
G23.707$-$00.198 & 33, 45    & $-$52.7 &  57 & 115 & 37 &  1.2 \\
G24.541$+$00.312 & $-$52, $-$15 & $-$15.5 & 112 & 52 & 48 &  2.4 \\
\\
\hline\hline
\multicolumn{7}{l}{$^*$ coordinates relative to the brightest spots
as listed in Table \ref{table:5}.}\\
\end{tabular}
\end{table}

We obtained reasonable fits for only two of the arched sources,
G23.707$-$00.198 and G24.541$+$00.312. The following best-fit model parameters are
listed in Table \ref{table:11}: the position of the vertex, the
opening angle of the cone (2$\theta$), the inclination angle between
the outflow axis and the z$-$axis ($\psi$), and the direction of the
x$-$axis, PA, which is the position angle of the outflow on the plane
of the sky.  A comparison between the modelled and observed data and a
sketch of the orientation of the outflow was presented previously in
Fig.~1 in Bartkiewicz et al. (\cite{bartkiewicz06}).  It is 
significant that the vertices of the cones calculated for both sources
coincide with infrared sources within the position uncertainties
(Sect.~\ref{sec:mir}). In addition, molecular line emission at
similar LSR velocities was reported towards both sources (Szymczak et
al. \cite{szymczak07}). As we mentioned previously, Araya et
al. (\cite{araya06}) imaged H$_2$CO maser emission at 4.8\,GHz towards
G24.541$+$00.312. All of these findings ensure that the outflow scenario is 
plausible for these two objects.

We tested the outflow model intensively on the four maser sources
associated with H\,{\small II} regions. However, we did not achieve a
reasonable fit that would place the vertex of the cone at the center 
of the radio continuum object, whilst reproducing the maser spot
kinematics, for any of these sources.

\subsection{Association with MIR emission}
\label{sec:mir}
Early work by Szymczak et al. (\cite{szymczak02}) showed that
$\sim$80\% of methanol masers have infrared counterparts in IRAS
and/or MSX catalogues. Since 
maser data with position accuracy as good or  better than these IR
catalogues (30\arcsec$-$5\arcsec) have become available, the fraction
of secure identifications has diminished (Pandian et
al.\,\cite{pandian07}). We can now attempt identifications with 
sub-arcsec  resolution data.

We used {\emph Spitzer} IRAC data to test the association between methanol
masers and MIR emission. Images at 3.6, 4.5, 5.8, and 8.0\,$\mu$m from
the GLIMPSE survey, all at 0\farcs6 resolution (Fazio et
al.\,\cite{fazio04}), were retrieved from the {\emph Spitzer}
archive\footnote{http://irsa.ipac.caltech.edu/applications/Cutouts/}
and compared with the radio data using AIPS.

   \begin{figure}
   \centering
   \includegraphics{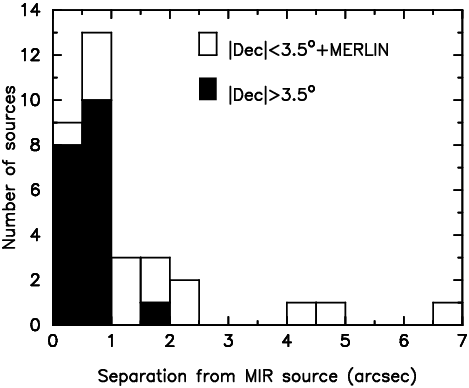}
   \caption{Histogram of number of the methanol masers as a function
            of angular separation from associated MIR sources at
            4.5\,$\mu$m. Bin size is 0\farcs5. The sources are divided
            into two groups with position measurements of high
            accuracy ($|$Dec$|>$3\fdg5) and lower accuracy
            ($|$Dec$|<$3\fdg5 plus those measured with MERLIN alone).}
   \label{fig11}
   \end{figure}

The angular separation between the brightest maser spot of each source
in our sample and the nearest MIR source at 4.5\,$\mu$m ($\Delta_{\rm
MIR}$) is given in Table \ref{table:5} and a histogram of the results
is shown in Fig. \ref{fig11}.  The average separation for the entire 
sample is 1\farcs18$\pm$0\farcs19 and the median value is 0\farcs89,
whereas, for the subsample of 19 objects at $|$Dec$|>$3\fdg5 with
coordinates measured using the EVN, the corresponding values are
0\farcs61$\pm$0\farcs09 and 0\farcs51, respectively. The remaining 14
targets with $|$Dec$|<$3\fdg5, or with positions derived from the
MERLIN data alone, show a larger $\Delta_{\rm MIR}$ with the mean and
median values of 1\farcs96$\pm$0\farcs35 and 1\farcs61, respectively. 
Since the images of these sources had poor  $uv$-plane coverage and hence
less accurate astrometry, we suggest that their higher estimates of
$\Delta_{\rm MIR}$ have a higher uncertainty and that most associations
are also likely to be genuine.

We conclude that the majority of maser sources coincide, within
$\la$1\arcsec\, with MIR emission, i.e., the maser emission from each
source falls within one \emph{Spitzer} pixel (nominal size $\sim$1\farcs2,
see Fazio et al.\,\cite{fazio04}) in the IRAC 4.5\,$\mu$m image.  This
strongly reinforces the finding by Cyganowski et
al. (\cite{cyganowski08}), who reported coincidences to within
$\la$5\arcsec\, for 46 out of 64 methanol sources with positions
measured using the ATCA.  Extended emission in the IRAC 4.5\,$\mu$m
band has been postulated to be a tracer of shocked molecular gas in
protostellar outflows (e.g., Davies et al.\,\cite{davis07}; Cyganowski
et al.\,\cite{cyganowski08} and references therein). This IRAC band
contains H$_2$ and CO lines that may be excited by shocks.  Using
shock models, Smith \& Rosen (\cite{smithrosen05}) predicted that H$_2$
emission in outflows is 5--14 times stronger in the 4.5\,$\mu$m band
than in the 3.6\,$\mu$m band. Cyganowski et al. (\cite{cyganowski08})
identified more than 300 objects with extended 4.5\,$\mu$m emission, 
which may relate to outflows in massive stars. Four maser sources from
our sample, G23.966$-$00.109, G35.793$-$00.175, G37.479$-$00.105, and
G39.100$+$00.491, were included in their catalogue.

In order to search for extended emission, we created images of the
4.5\,$\mu$m$-$3.6\,$\mu$m excess by subtracting the 3.6\,$\mu$m
image of each of our sources from its 4.5\,$\mu$m counterpart, and
compared these with the  4.5\,$\mu$m image.
All the sources have  extended 4.5\,$\mu$m emission. 
Figure \ref{fig12} shows the
4.5\,$\mu$m$-$3.6\,$\mu$m excess superimposed on the  4.5\,$\mu$m image
for selected sources. Source G21.407$-$00.254
illustrates how the maser emission located precisely inside the brightest
pixel of the 4.5\,$\mu$m$-$3.6\,$\mu$m image outside the 4.5\,$\mu$m peak.  
There is also evidence that maser clusters in at least three sources
are associated with 4.5\,$\mu$m emission excess.  The image at the
position of G24.148$-$00.009 shows that the maser 
coincides exactly with maxima of both the 4.5\,$\mu$m$-$3.6\,$\mu$m excess 
and of the 4.5\,$\mu$m emission. Weak, extended 4.5\,$\mu$m emission is also
seen at the edges of two neighbouring sources and in diffuse lanes.
The maser G37.598$+$00.425 coincides with
a maximum in a very asymmetric distribution of 4.5\,$\mu$m$-$3.6\,$\mu$m
emission excess that is displaced by $\sim$1\farcs5 from a peak of
4.5\,$\mu$m emission, implying  that the methanol maser and the
excess in the 4.5\,$\mu$m IRAC band are very strongly related. 
Similar evidence  is provided by  G38.203$-$00.067, where the maser is
offset by more than 8\arcsec\, from a bright MIR object. The maser is
$\sim$1\farcs7 away from a weak bump in a large lane of diffuse
4.5\,$\mu$m$-$3.6\,$\mu$m emission excess. Inspection of IRAC images
for other bands suggests that the maser is probably associated with a faint MIR
object.

All the counterparts in the sample have MIR properties typical of
embedded young massive objects (e.g., Kumar et al. 2007) associated
with the methanol masers (Ellingsen 2006). We defer a detailed
discussion of the MIR spectral indices of individual objects, because several
bright sources (e.g., G23.389$+$00.185, G23.657$-$00.127, and 
G24.634$-$00.324) are saturated in the IRAC images, while for others
only upper limits to the 3.6\,$\mu$m flux can be derived
(e.g., G23.966$-$00.109, G37.030$-$00.039, and G38.203$-$00.067).  We 
note only, that the MIR objects associated with methanol masers that have a
ring-like morphology have much stronger emission at 8\,$\mu$m, and all
the objects that are saturated in the IRAC images have regular maser
structures.  In contrast, no object saturated in the IRAC images has a
complex maser morphology. This suggests that the MIR counterparts of
masers with less regular structures are deeply embedded massive
stars that are younger than the counterparts of those with more
regular, ring-like maser structures. One can speculate that in those
younger objects the methanol maser originates in a limited number of
confined regions, whilst more regular structures emerge during later
evolutionary stages.

In summary, we have found strong evidence of a close coincidence of
6.7\,GHz methanol masers with 4.5\,$\mu$m emission excess. This
provides a firm argument that methanol emission originates in those inner
parts of an outflow or disc/torus where the molecular gas is shocked.

This hypothesis is strongly supported by the kinematic model that can be
successfully applied to the ring masers in our sample, which involved a rotating
and expanding disc wherein expansion dominates the velocity field. Moreover, 
a fraction of masers seem to fit the model of a bipolar jet or that of 
a shock front colliding with the surrounding molecular material.

\onlfig{6}{
   \begin{figure*}
   \centering
   \includegraphics{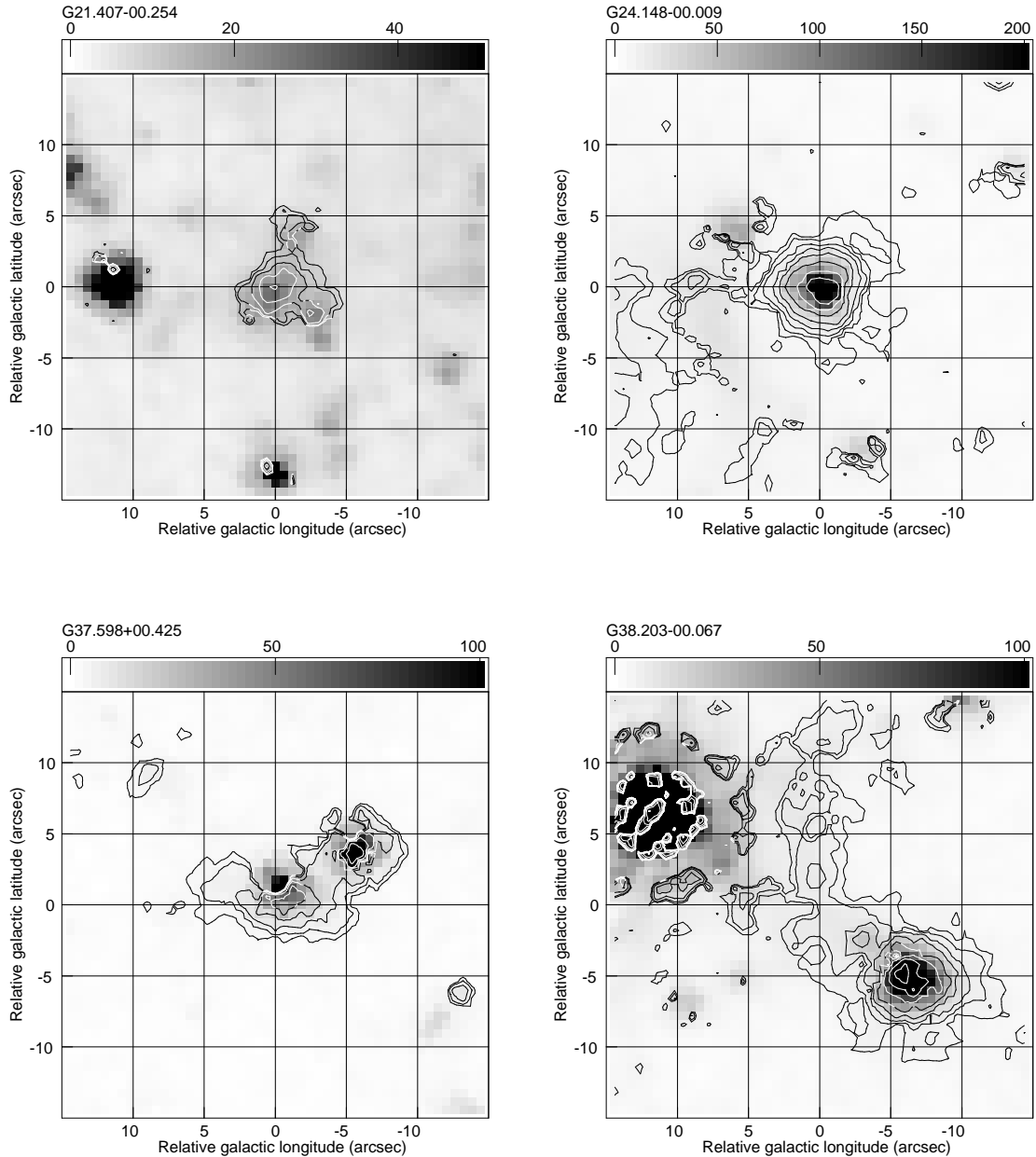}
   \caption{{\emph Spitzer} IRAC 4.5\,$\mu$m (grey) and
   4.5\,$\mu$m$-$3.6\,$\mu$m (contour) images centered on selected
   methanol sources
   (at coordinates taken from Table \ref{table:5}). The grey scale in
   MJy\,sr$^{-1}$ is indicated in the horizontal bar and contour
   levels are 1\,MJy\,sr$^{-1}\times$(1, 2, 4, 8, 16, 32, 64,
   128). The methanol maser {\bf G21.407$-$00.254} coincides exactly 
   with the brightest pixel in the 4.5\,$\mu$m$-$3.6\,$\mu$m image, at
   the map origin, which is offset from the 4.5\,$\mu$m peak. There
   are at least two other nearby MIR sources with 4.5\,$\mu$m emission
   excesses. Note the absence of further sources with extended
   4.5\,$\mu$m$-$3.6\,$\mu$m emission excess in this 30\arcsec
   $\times$30\arcsec\, field.  The maser source {\bf G24.148$-$00.009}
   coincides with the peaks of emission both at 4.5\,$\mu$m and for the
   4.5\,$\mu$m$-$3.6\,$\mu$m excess.  The maser {\bf G37.598$+$00.425}
   is offset from the brightest 4.5\,$\mu$m emission. Note the highly
   asymmetric morphology of the extended 4.5\,$\mu$m$-$3.6\,$\mu$m excess
    emission and the coincidence of the maser emission with a maximum of
   this excess. In the case of {\bf G38.203$-$00.067}, the maser
   emission does not coincide with either of the two brightest sources
   but is close ($\sim$1\farcs7) to a weak bump of
   4.5\,$\mu$m$-$3.6\,$\mu$m excess located in a very extended arc
   ($\sim$25\arcsec) of diffuse 4.5\,$\mu$m emission. Further
   inspection of IRAC images at 5.8\,$\mu$m suggests that this bump
   could be a weak MIR object.}
   \label{fig12}
   \end{figure*}
}

\section{Conclusions} 
We have completed a 6.7\,GHz methanol line imaging survey of 
 33 maser sources in the Galactic plane. High quality EVN images
were obtained for 31 targets showing their mas-scale structures, from
which we derived the absolute positions of 29 sources with a few mas accuracy. 
In most cases, the masers exhibit complex structures. The
observed morphologies can be divided into five groups: simple, linear,
ring-like, complex, arched, and pair. It is surprising that about 29\%
of the sources exhibit ring-like distributions of maser spots that 
were not apparent in previous VLBI surveys, which were 
less sensitive and concentrated on brighter masers. 
We find that many of the other ordered structures, notably linear and 
arched, can be interpreted as originating from the interaction 
 between collimated or biconical outflows and the surrounding medium.

A simultaneous survey at high angular resolution of continuum emission
at 8.4\,GHz towards the maser targets revealed that only 16\% of the
methanol masers appear to be physically related to H\,{\small II}
regions. In three cases, the maser coincides with a peak in a compact
and relatively weak continuum object. These are believed to be young
UC H\,{\small II} regions. In general, these results seem to imply
that 6.7\,GHz methanol masers appear before the ionised region is
detectable at cm wavelengths. The hypothesis that these masers are
possible associated with HC H\,{\small II} regions needs to be verified.

We used the \emph{Spitzer} GLIMPSE survey to demonstrate that the
majority of methanol masers are closely associated with MIR
emission. Analyzis of the MIR counterparts suggests that masers with a
regular, ring-like morphology are associated with more evolved proto-
or young stars, relative to the counterparts of masers with more
complex, irregular structures. Moreover, both the kinematics of the
masers with a ring-like distribution and the characteristics
of associated MIR emission are consistent with a scenario in which the
methanol emission is produced by shocked material associated 
with a disk or torus, possibly from an interaction with the outflow.

\begin{acknowledgements}
Our special acknowledgements to Dr.\,Peter Thomasson at Jodrell Bank
Observatory and to Dr.\,Bob Campbell at JIVE for detailed support in
many stages of this project. We also thank  Dr.\,Riccardo Cesaroni,
Dr.\,Luca Moscadelli, Dr.\,Lucero Uscanga, Dr.\,Krzysztof
Go\'zdziewski and Kalle Torstensson for useful discussions.  This work
was benefited from the Polish MNiI grant 1P03D02729 and from research
funding from the EC 6th Framework Programme. \\ The European VLBI
Network (EVN) is a joint facility of European, Chinese, South African
and other radio astronomy institutes funded by their national research
councils.  MERLIN is a National Facility operated by the University of
Manchester at Jodrell Bank Observatory on behalf of STFC.  The Very
Large Array (VLA) of the National Radio Astronomy Observatory is a
facility of the National Science Foundation operated under cooperative
agreement by Associated Universities, Inc.  This research has made use
of the NASA/IPAC Infrared Science Archive, which is operated by the
Jet Propulsion Laboratory, California Institute of Technology, under
contract with the National Aeronautics and Space Administration.

\end{acknowledgements}

\Online

\clearpage
\onecolumn
\setcounter{table}{5}
\begin{table}
\caption{Maser clusters towards 31 sources.  If a spectrum of a maser 
cluster does not show a Gaussian profile we enter the sign $-$ in the 
Cols. of FWHM and S$_{\rm amp}$.}
\label{table:6}      
\centering          
\begin{supertabular}{rrrccc}     
\hline\hline       
\\
V$_{\rm LSR}$ & $\Delta$RA & $\Delta$Dec & S$_{\rm p}$& FWHM & S$_{\rm amp}$\\
(km\,s$^{-1}$) & (mas)     &  (mas)      & (mJy\,beam$^{-1}$)&
(km\,s$^{-1}$)& (mJy\,beam$^{-1}$)\\
\hline
\\
\multicolumn{6}{l}{G21.407$-$00.254}\\
89.040&    0.0000   &  0.0000 & 2762& 0.32&2672\\
89.303& $-$0.0192   &  9.7250 & 1426& 0.35&1513\\
90.094& $-$8.9947   & $-$26.3260& 307 & 0.18&308\\
91.148& $-$136.3023 &  5.0420 & 891 & 0.26&888\\
91.412& $-$13.1560  & $-$23.6830& 117 & $-$&$-$\\
91.587& $-$27.3535  &  1.5320 & 422 & 0.26&420\\
\multicolumn{6}{l}{G22.335$-$00.155}\\
35.626 &  0.0000   &  0.0000 & 1714 &0.32&1642\\
38.174 &$-$10.9782 &  $-$47.4060 &125 & 0.53&125\\
\multicolumn{6}{l}{G22.357$+$00.066}\\
79.708 & 0.0000   &  0.0000 & 10544& 0.37& 11350\\
80.235 & 1.0074   & $-$6.0100 & 10067& 0.47& 10190\\
81.114 & 2.8964   & $-$18.3490& 1422 & 0.48& 1497\\
81.465 & 155.0868 & $-$46.3230& 492  & $-$   &$-$\\
83.574 & 139.2789 & $-$18.6790& 309  & $-$   &$-$\\
84.101 & 141.5278 &  66.5690& 156  & 0.58& 158\\
84.804 & 132.0327 &  75.7690& 176  & $-$   &$-$\\
88.143 & 55.7575  & 208.5390& 602  & 0.27& 636\\
88.494 & 68.1772  & 277.2010& 250  & $-$   &$-$\\
88.670 & 56.6464  & 207.5440& 476  & $-$   &$-$\\
\multicolumn{6}{l}{G23.207$-$00.377}\\
72.553 &$-$101.3966 & $-$64.2970&  141 & 0.31& 140\\
73.959 &$-$32.5775  & $-$17.5310&  194 & 0.28& 190\\
75.190 & 0.5751   &  9.9430 &  226 & 0.23& 223\\
75.014 &$-$25.7012  & $-$1.9800 &  421 & 0.83& 374\\
75.453 &$-$269.6437 &  26.1920&  893 & 0.34& 946\\
75.629 & 57.3581  &  41.8210&  647 & 0.34& 683\\
76.069 &$-$30.8269  & $-$11.6140&  229 & 0.42& 228\\
76.596 &$-$11.6581  &  3.1990 &  4748& 0.43& 4761\\
77.123 & 0.0000   &  0.0000 &  9292& 0.34& 9468\\
77.475 & 29.4751  & $-$4.2880 &  3387& 0.46& 3463\\
77.739 & 11.0845  &  4.9910 &  3096& 0.47& 2852\\
78.793 & 53.0729  & $-$33.4270&  541 & 0.44& 563\\
78.881 &$-$217.1237 &  30.9130&  286 & 0.28& 293\\
78.969 & 47.4388  & $-$30.9230&  577 & $-$   &$-$\\
79.233 & 50.4581  & $-$32.3460&  1319& 0.38& 1412\\
79.496 &$-$135.9039 & $-$191.481&0 277 & 0.21& 294\\
79.672 &$-$113.8403 & $-$117.084&0 213 & 0.41& 221 \\
79.936 &$-$57.8547  & $-$71.4830&  315 & 0.29& 303 \\
80.727 &$-$163.9145 & $-$147.143&0 7092& 0.36& 6989\\
81.869 &$-$136.2582 & $-$139.260&0 8967& 0.37& 8838\\
81.869 &$-$141.0800 & $-$140.772&0 5874& 0.60& 5918\\
82.045 &$-$157.2072 & $-$145.562&0 1104& 0.25& 1119\\
82.836 &$-$151.9599 & $-$147.620&0 1190& 0.43& 1012\\
84.858 &$-$47.3987  & $-$112.077&0 791 & 0.56& 679\\
\multicolumn{6}{l}{G23.389$+$00.185}\\
72.641 &$-$78.2268  & 32.6440 &  6398&  0.46& 6279\\
73.784 & 52.4158  & 32.4260 &  9326&  0.47& 8831\\
74.399 & 46.8761  & 38.8880 &  2978&  0.41& 2715\\
74.575 &$-$43.2789  & 31.9400 &  4700&  0.28& 4730 \\
74.750 & 43.4540  & 63.9750 &  311 &  $-$   &    $-$  \\
74.750 & 41.0708  & 14.0150 &  267 &  $-$   &    $-$  \\
74.838 & 31.5347  & 70.8960 &  19723& 0.71& 21320\\
74.926 & 41.8691  & 49.9560 &  1272&  0.17& 1232 \\
75.453 &  0.0000  & 0.0000  &  21554& 0.26& 21587\\
75.541 &$-$57.2121  & 153.9020&  397 &  $-$   &   $-$ \\
\hline
\end{supertabular}
\end{table}
\begin{table}
\centering
\addtocounter{table}{-1}
\caption{ continued.}
\begin{supertabular}{rrrccc}
\hline\hline       
\\
V$_{\rm LSR}$ & $\Delta$RA & $\Delta$Dec & S$_{\rm p}$& FWHM & S$_{\rm amp}$\\
(km\,s$^{-1}$) & (mas)     &  (mas)      & (mJy\,beam$^{-1}$)&
(km\,s$^{-1}$)& (mJy\,beam$^{-1}$)\\
\hline
\\
75.805 &$-$64.5014  & 143.6490&  3708&  0.32& 3768 \\
76.069 & 78.5948  & 144.9390&  6571&  0.37& 6686 \\
76.332 &$-$93.2091  & 154.8360&  4516&  0.62& 4683 \\
76.596 &$-$70.0292  & 145.2360&  1072&  0.55& 1086 \\
76.948 & 20.9791  & 75.4450 &  721 &  0.31& 727  \\
77.035 &$-$69.1893  & 147.1100&  4860&  0.28& 4795 \\
77.475 &$-$73.8030  & 150.8570&  234 &  0.19& 234  \\
77.563 &$-$31.8834  & 157.9720&  485 &  0.24& 470\\
\multicolumn{6}{l}{G23.657$-$00.127}\\
77.563 & $-$5.0421  &  229.8720&  816 & 0.42& 754  \\
77.563 & $-$22.2581 &  232.7000&  333 & 0.25& 355  \\
78.266 & $-$71.1275 &  228.1640&  1059& 0.52& 932  \\
78.881 &  54.5928 &  171.6190&  336 & 0.50& 305  \\
79.584 &  105.3428&  117.0720&  162 & 0.75& 170  \\
80.024 & $-$124.7437&  212.4980&  1595& 0.23& 1607 \\
80.112 & $-$39.3568 &  213.5540&  977 & 0.32& 938  \\
80.639 &  21.7712 & $-$13.2960 &  850 & 0.49& 691\\
80.727 &  14.3113 & $-$66.4860 &  375 & 0.20& 391\\
80.990 & $-$156.9775& $-$5.2880  &  124 & 0.25& 135\\
81.166 &  2.9967  & $-$47.4670 &  362 & 0.30& 370\\
81.166 &  26.3472 & $-$70.2780 &  68  & 0.55& 77 \\
81.518 &  7.2432  & $-$3.7020  &  588 & 0.42& 564\\
81.781 &  46.3136 &  156.5250&  126 & 0.39& 117\\
81.869 &  10.2711 &  10.8770 &  806 & 0.30& 820 \\
81.869 &  13.2471 & $-$13.0970 &  90  & 0.19& 92  \\
82.133 & $-$1.5481  & $-$43.7250 &  132 & 0.31& 124 \\
82.309 &  44.5072 &  156.8130&  393 & 0.28& 391 \\
82.573 &  0.0000  &  0.0000  &  3623& 0.48& 3401\\
82.836 &  41.5298 &  162.2880&  168 & 0.29& 174 \\
83.100 & $-$210.9484&  11.3450 &  182 & 0.24& 196 \\
83.276 & $-$47.8646 & $-$35.5030 &  1018& 0.33& 1036\\
83.539 &  69.2069 &  83.7510 &  80  & 0.27& 81  \\
83.979 &  0.7674  &  18.7780 &  143 & 0.62& 125 \\
83.979 & $-$52.8072 & $-$32.5950 &  557 & 0.58& 531 \\
84.243 & $-$133.7606& $-$4.1400  &  358 & 0.27& 373 \\
84.858 &  45.6887 &  140.0160&  548 & 0.44& 552 \\
86.264 & $-$17.8008 & $-$15.6640 &  797 & 0.37& 766 \\
86.791 & $-$19.7200 & $-$18.7010 &  746 & 0.37& 702\\
87.231 & $-$16.7410 & $-$26.9680 &  96  & 0.24& 96\\
87.670 & $-$20.3523 & $-$29.5260 &  280 & 0.26& 272\\
\multicolumn{6}{l}{G23.707$-$00.198}\\
58.578 &  20.4283 &  18.1540 &  377 & $-$   &\\
71.410 & $-$101.2879&  0.5490  &  200 & 0.19& 207\\
71.762 & $-$105.4053& $-$5.6330  &  128 & 0.19& 141\\
72.993 & $-$10.2000 &  57.1700 &  667 & 0.23& 707\\
73.344 & $-$1.6921  &  33.7300 &  92  & 0.40& 95 \\
73.608 &  0.3488  &  27.4820 &  100 & 0.22& 102\\
74.223 & $-$3.2461  &  38.1980 &  672 & 0.38& 683\\
74.575 &  0.3933  &  23.6850 &  1845& 0.26& 1756\\
75.102 & $-$0.0089  &  29.0830 &  404 & 0.47& 416  \\
75.190 &  16.6285 & $-$9.9600  &  306 & 0.52& 340  \\
75.629 &  0.3577  &  26.8210 &  249 & 0.50& 260  \\
76.157 &  0.9722  &  19.8400 &  3319& 0.46& 3297 \\
76.596 &  0.9589  &  17.6580 &  2054& 0.83& 2033 \\
77.299 &  0.8772  &  16.2390 &  3473& 0.29& 3625 \\
78.266 & $-$14.6856 & $-$14.3980 &  219 & 0.31& 219  \\
78.530 & $-$13.7935 & $-$14.4180 &  358 & 0.32& 366  \\
\hline
\end{supertabular}
\end{table}
\begin{table}
\centering
\addtocounter{table}{-1}
\caption{ continued.}
\begin{supertabular}{rrrccc}
\hline\hline       
\\
V$_{\rm LSR}$ & $\Delta$RA & $\Delta$Dec & S$_{\rm p}$& FWHM & S$_{\rm amp}$\\
(km\,s$^{-1}$) & (mas)     &  (mas)      & (mJy\,beam$^{-1}$)&
(km\,s$^{-1}$)& (mJy\,beam$^{-1}$)\\
\hline
\\
77.651 & $-$0.4334  &  3.7530  &  2234& 0.70& 2239 \\
78.090 & $-$0.5982  &  1.0340  &  2144& 0.95& 2099 \\
79.145 &  0.0000  &  0.0000  &  6059& 0.79& 5368 \\
79.496 & $-$17.3246 & $-$13.9080 &  117 & 0.61& 109  \\
80.112 & $-$14.3234 & $-$14.3790 &  202 & 0.36& 200  \\
80.903 & $-$4.2421  & $-$8.0300  &  142 & 0.48& 129  \\
81.078 & $-$3.9200  & $-$8.6280  &  159 & 0.56& 153\\
\multicolumn{6}{l}{G23.966$-$00.109}\\
67.428 & 21.9498  & $-$23.0050 &387 & 0.30& 369\\
68.219 & 23.3000  & $-$22.9600 &431 & 0.41& 404\\
70.942 & 0.0000   &  0.0000  &5487& 0.41& 5390\\
\multicolumn{6}{l}{G24.148$-$00.009}\\
17.441 & 4.3548   & $-$21.7190 &1048& 0.20& 1048\\
17.529 & 0.1980   & $-$0.0970  &2459& 0.51& 2493\\
17.792 & 0.0000   &  0.0000  &3648& 0.40& 3702\\
18.407 & 0.6149   &  5.9860  &240 & 0.33& 235\\
\multicolumn{6}{l}{G24.541$+$00.312}\\
103.754 & $-$35.7935 &  59.5820 & 233  & 0.23& 233  \\
105.688 &  0.0000  &  0.0000  & 7753 & 0.34& 7453 \\
106.215 & $-$38.6463 &  37.3600 & 316  & 0.29& 278  \\
107.973 & $-$10.8681 &  7.8090  & 238  & 0.33& 239  \\
106.479 & $-$5.5553  &  2.9680  & 2788 & 0.49& 2554 \\
107.270 &  0.0253  &  1.0010  & 2700 & 0.47& 2677 \\
107.533 & $-$1.8840  &  2.4950  & 2005 & 0.38& 2052 \\
108.324 &  94.8606 &  14.3270 & 106  & 0.32& 108  \\
108.588 & $-$26.0490 &  13.4890 & 638  & 0.37& 534  \\
109.467 & $-$39.7491 &  16.3690 & 339  & 0.57& 339  \\
109.643 & $-$39.9396 &  16.9440 & 407  & 0.38& 413  \\
110.082 & $-$41.4218 &  18.0810 & 728  & 0.39& 742  \\
\multicolumn{6}{l}{G24.634$-$00.324}\\            
34.725 &$-$18.6698  &  25.4370 &457  & $-$  &    $-$ \\
35.428 & 0.0000   &  0.0000  &3027 &0.29& 3544    \\
43.862 &$-$62.2684  &  36.4860 &2370 &0.56& 2322    \\
45.795 &$-$41.9147  &  45.7150 &257  &0.41& 257     \\
46.322 &$-$43.8591  &  45.1230 &312  &0.38& 313     \\
47.904 &$-$42.4775  &  45.4230 &136  &0.54& 136\\
\multicolumn{6}{l}{G25.411$+$00.105}\\
93.765 & 34.1096  &  111.2790 & 142 & $-$   &    $-$ \\
94.117 & 109.0956 & $-$19.4400  & 82  & $-$   &  $-$   \\
94.820 & 115.8608 &  101.1940 & 847 & $-$   &  $-$   \\
94.644 &$-$16.7363  &  86.3980  & 2374& 0.51& 2345\\
94.820 &$-$3.5683   & $-$20.6650  & 1025& 0.50& 1037\\
96.928 &$-$3.5146   &  123.8050 & 323 & $-$   &  $-$   \\
96.928 & 3.2029   & $-$5.4840   & 2251& 0.42& 2241\\
97.104 &$-$46.6480  &  16.7050  & 2468& 0.28& 2504\\
97.280 & 0.0000   &  0.0000   & 3433& 0.56& 3690\\
98.861 & 176.2362 &  50.0100  & 139 & $-$   &   $-$  \\
\multicolumn{6}{l}{G26.598$-$00.024}\\
22.952&  2.1488  &  $-$0.1130   & 182 & 0.45& 183  \\
24.182&  0.0000  &   0.0000   & 3043& 0.34& 3239 \\
24.709&  84.5157 &  $-$169.5700 & 1368& 0.44& 1276 \\
25.061&  335.4955&  $-$131.1280 & 714 & 0.51& 671  \\
25.588&  78.8583 &  $-$159.7380 & 76  & $-$   &   $-$   \\
25.939&  338.8875&  $-$124.8040 & 97  & 0.94& 99\\
\hline
\end{supertabular}
\end{table}
\begin{table}
\centering
\addtocounter{table}{-1}
\caption{ continued.}
\begin{supertabular}{rrrccc}
\hline\hline       
\\
V$_{\rm LSR}$ & $\Delta$RA & $\Delta$Dec & S$_{\rm p}$& FWHM & S$_{\rm amp}$\\
(km\,s$^{-1}$) & (mas)     &  (mas)      & (mJy\,beam$^{-1}$)&
(km\,s$^{-1}$)& (mJy\,beam$^{-1}$)\\
\hline
\\
\multicolumn{6}{l}{G27.221$+$00.136}\\
105.424 & $-$11.6570 &  77.8590 & 253  & 0.42& 237  \\
106.391 & $-$11.2849 &  77.8360 & 94   & 0.59& 91   \\
107.182 & $-$33.2614 &  55.4880 & 245  & 0.42& 232  \\
109.731 & $-$33.8577 &  32.9640 & 1239 & 0.26& 1193 \\
110.258 & $-$34.2955 &  34.2190 & 1182 & 0.43& 1179 \\
111.225 & $-$33.0149 &  35.1860 & 467  & 0.30& 466  \\
112.280 & $-$19.4363 &  27.0680 & 563  & 0.35& 538  \\
112.719 & $-$18.6697 &  27.1140 & 906  & 0.29& 917  \\
114.828 & $-$20.1700 &  27.8700 & 166  & 0.25& 169  \\
115.268 & $-$18.4171 &  23.1670 & 150  & 0.39& 152  \\
115.444 & $-$19.0119 &  24.4340 & 158  & $-$   &  $-$ \\
115.707 & $-$24.1941 &  47.1650 & 236  & 0.31& 225 \\
116.586 & $-$24.7798 &  7.8580  & 437  & 0.23& 457 \\
117.114 & $-$53.1908 &  8.6200  & 6519 & 0.33& 6591\\
117.729 &  8.9344  & $-$12.1430 & 448  & 0.23& 469 \\
117.817 & $-$54.3145 &  7.6670  & 5642 & 0.55& 5621 \\
118.168 & $-$9.3558  & $-$11.4900 & 1991 & 0.33& 1990 \\
118.432 & $-$55.3082 &  7.0300  & 4831 & 0.35& 4755 \\
118.783 &  0.0000  &  0.0000  & 12535& 0.39& 12021\\
118.959 & $-$8.6878  &  13.1750 & 765  & 0.67& 801  \\
119.223 & $-$70.7368 & $-$6.8450  & 209  & $-$   &  $-$    \\
119.311 & $-$9.2317  & $-$5.7490  & 232  & 0.33& 224  \\
119.311 & $-$34.3642 &  4.9240  & 135  & 0.51& 111  \\
119.750 & $-$9.4066  &  11.3470 & 101  & 0.38& 103  \\
120.365 & $-$15.4465 & $-$18.4780 & 536  & 0.23& 536  \\
120.190 & $-$20.1998 &  41.8670 & 1339 & 0.40& 1367\\
121.069 & $-$17.4294 &  44.0510 & 218  & 0.29& 218\\
\multicolumn{6}{l}{G28.817$+$00.365}\\
87.735 &$-$73.2105 &  $-$56.0600 &  325 & 0.51 &331   \\
88.087 &$-$63.9892 &  $-$72.9580 &  183 & $-$    &  $-$ \\
90.723 & 0.0000  &  0.0000   &  3137& 0.36 &3211  \\
91.074 & 2.8641  &  17.4230  &  1440& 0.62 &1440  \\
91.250 & 2.7787  &  3.5790   &  896 & 0.68 &933   \\
91.601 & 4.0115  &  20.0900  &  2511& 0.31 &2755  \\
92.129 &$-$3.1337  &  10.3800  &  743 & 0.47 &761\\
92.656 &$-$2.7143  &  26.7830  &  2116& 0.38 &2103\\
\multicolumn{6}{l}{G30.318$+$00.070}\\
35.192& $-$23.8385 &  22.6390 &   300 & 0.53& 303\\
36.070&  0.0000  &  0.0000  &   514 & 0.68& 528\\
36.949&  5.6770  & $-$5.0420  &   144 & 0.70& 144\\
\multicolumn{6}{l}{G30.400$-$00.296}\\
98.104 &  4.4898  &$-$11.8250 & 2139& 0.31 &2151 \\
98.455 &  0.0000  & 0.0000  & 2765& 0.32 &3129 \\
100.388& $-$31.0281 &$-$45.2180 & 1581& 0.29 &1603 \\
101.618& $-$27.7091 & 85.7960 & 331 & 0.72 &326  \\
103.727& $-$19.1733 & 86.2030 & 1914& 0.41 &1923 \\
103.727&  67.4799 & 134.7790& 497 & 0.28 &503  \\
104.430&  66.4200 & 135.2860& 260 & 0.43 &267\\
\hline
\end{supertabular}
\end{table}
\begin{table}
\centering
\addtocounter{table}{-1}
\caption{ continued.}
\begin{supertabular}{rrrccc}
\hline\hline       
\\
V$_{\rm LSR}$ & $\Delta$RA & $\Delta$Dec & S$_{\rm p}$& FWHM & S$_{\rm amp}$\\
(km\,s$^{-1}$) & (mas)     &  (mas)      & (mJy\,beam$^{-1}$)&
(km\,s$^{-1}$)& (mJy\,beam$^{-1}$)\\
\hline
\\
\multicolumn{6}{l}{G31.047$+$00.356}\\
81.058 & $-$59.6159& $-$28.5990&  277 & $-$   &  $-$ \\
78.070 & $-$0.9674 &  7.7850 &  281 & 0.82& 267  \\
79.125 & $-$1.2973 &  3.2920 &  1169& 0.73& 1180 \\
80.179 &  0.4664 &  1.5320 &  544 & 0.55& 555  \\
80.882 &  0.0000 &  0.0000 &  1989& 0.55& 1960 \\
81.058 & $-$0.3989 & $-$1.5250 &  1563& 0.55& 1568 \\
82.991 & $-$47.8577& $-$6.9740 &  617 & 0.28& 651\\
84.045 & $-$25.8591&  9.6530 &  101 & $-$   &$-$\\
\multicolumn{6}{l}{G31.581$+$00.077}\\
95.643 &  0.0000  & 0.0000  & 2722& 0.35& 2891 \\
97.752 & $-$60.9387 & 57.6080 & 349 & $-$   & $-$   \\
98.104 &  37.3458 &$-$7.0480  & 809 & 0.31& 814  \\
98.455 &  32.2616 &$-$3.7570  & 376 & 0.39& 386  \\
98.631 &  149.7072& 94.0260 & 493 & $-$   &   $-$\\
98.806 & $-$61.9691 & 79.9160 & 292 & 0.41& 321  \\
98.806 &  149.8767& 93.2150 & 2045& 0.68& 2039\\
99.509 &  158.1329& 69.4150 & 358 & $-$   &$-$\\
99.685 &  151.6794& 56.5090 & 1099& 0.39& 1223\\
\multicolumn{6}{l}{G32.992$+$00.034}\\
89.724 &  13.3545 &$-$44.5810 & 321 & $-$   & $-$      \\
90.339 & $-$51.5775 & 49.8690 & 396 & $-$   &    $-$   \\
90.602 &  14.7555 &$-$34.4810 & 1967& 0.40& 1953  \\
90.954 &  26.7960 & 1.1600  & 4650& 0.32& 4838  \\
91.393 & $-$4.7370  &$-$6.0210  & 493 & 0.50& 505   \\
91.481 &  28.3665 & 0.9760  & 1856& 0.50& 1867  \\
91.656 & $-$16.1355 & 37.9820 & 465 & $-$   &   $-$\\
91.832 &  0.0000  & 0.0000  & 6212& 0.35& 6786 \\
92.184 &  34.7340 &$-$14.9630 & 394 & 0.28& 394  \\
92.447 &  35.5635 &$-$14.7320 & 163 & $-$   &  $-$ \\
92.711 &  24.9990 &$-$19.6750 & 199 & $-$   &  $-$ \\
94.204 &  51.0915 &$-$1.6040  & 600 & 0.26& 574  \\
94.731 &  49.8900 &$-$4.5590  & 445 & 0.22& 446  \\
\multicolumn{6}{l}{G33.641$-$00.228}\\         
58.840 &  0.0000  & 0.0000  & 28300& 0.31 &29569 \\
59.540 & $-$9.5996  &$-$12.3000 & 4105 & 0.29 &4259  \\
59.800 & $-$14.3994 &$-$20.1000 & 12260& 0.22 &12665 \\
60.330 &  69.5970 &$-$15.9000 & 20402& 0.30 &21448 \\
60.420 &  61.9473 &$-$25.9000 & 2280 & 0.50 &2688  \\
60.510 &  60.8973 &$-$27.0000 & 1621 & $-$    &$-$   \\
60.860 &  13.3494 &$-$28.6000 & 1672 & 0.30 &1716  \\
60.940 &  65.3971 &$-$14.6000 & 10447& 0.35 &10558 \\
61.120 &  76.3467 &$-$2.3000  & 356  & 0.35 &372   \\
61.300 &  56.2475 &$-$17.0000 & 207  & 0.26 &213   \\
61.300 &  128.3944& 55.9000 & 176  & $-$    &  $-$ \\
61.820 &  82.1964 & 7.6000  & 365  & 0.40 &358   \\
62.260 &  122.5447& 63.5000 & 893  & 0.56 &812  \\
62.180 &  111.2951& 50.0000 & 109  & $-$    &  $-$\\
62.440 &  42.5981 &$-$18.2000 & 5440 & 0.31 &5758 \\
62.700 &  42.5981 &$-$17.8000 & 20690& 0.35 &20872\\
62.970 &  42.5981 &$-$16.6000 & 9325 & 0.48 &9620 \\
63.140 &  30.4487 &$-$19.7000 & 1090 & 0.30 &1124 \\
\multicolumn{6}{l}{G33.980$-$00.019}\\
58.886 &  0.0000 &  0.0000  & 3782& 0.33 &4057  \\
59.501 &  5.1225 &  0.1130  & 434 & 0.30 &454   \\
60.292 & $-$50.1990& $-$23.9650 & 315 & 0.45 &290   \\
\hline
\end{supertabular}
\end{table}
\begin{table}
\centering
\addtocounter{table}{-1}
\caption{ continued.}
\begin{supertabular}{rrrccc}
\hline\hline       
\\
V$_{\rm LSR}$ & $\Delta$RA & $\Delta$Dec & S$_{\rm p}$& FWHM & S$_{\rm amp}$\\
(km\,s$^{-1}$) & (mas)     &  (mas)      & (mJy\,beam$^{-1}$)&
(km\,s$^{-1}$)& (mJy\,beam$^{-1}$)\\
\hline
\\
60.555 & $-$29.6685&  1.7760  & 254 & 0.29 &258   \\
61.258 & $-$49.1925& $-$16.8830 & 784 & 0.37 &797   \\
61.609 &  30.1755& $-$5.5850  & 515 & 0.60 &469   \\
62.312 &  30.5595& $-$3.2980  & 276 & 0.40 &248 \\
64.157 &  16.2345& $-$27.9140 & 497 & 0.50 &471 \\
65.387 & $-$32.4450& $-$40.9110 & 236 & 0.25 &237 \\
\multicolumn{6}{l}{G34.751$-$00.093}\\        
50.627 &  28.8631& $-$27.8740&  224 & 0.59& 215 \\
51.769 & $-$10.8418& $-$3.5670 &  718 & 0.46& 702 \\
52.736 &  0.0000 &  0.0000 &  1954& 0.68& 1817\\
53.175 & $-$3.5980 & $-$0.9420 &  1763& 0.44& 1845\\
\multicolumn{6}{l}{G35.793$-$00.175}\\        
60.060 & $-$2.4000 & $-$0.8600 &  889 & 0.27&  952\\
60.680 &  0.0000 &  0.0000 &  9702& 0.62& 9616\\
61.300 &  1.9500 &  0.8100 &  6338& 0.48& 6327\\
61.820 &  3.0000 &  1.2100 &  5960& 0.75& 5927\\
62.350 &  4.3500 &  2.2100 &  1225& 0.60& 1347\\
\multicolumn{6}{l}{G36.115$+$00.552}\\
70.395 & $-$1072.0323& 421.7720& 10779& 0.38& 9795\\
70.834 & $-$1075.7487& 431.4520& 254  & 0.33& 261 \\
71.186 & $-$5.1454   & 0.8900  & 1066 & 0.40& 1113\\
71.449 & $-$1068.0478& 420.0420& 617  & 0.35& 637 \\
71.801 & $-$1062.5353& 420.6660& 2921 & 0.28& 2869\\
71.977 & $-$24.9857  &$-$12.5080 & 1386 & 0.29& 1398 \\
72.416 & $-$1064.1711& 421.8460& 2973 & 0.63& 2681 \\
72.504 & $-$1168.7980& 300.8850& 506  & $-$   & $-$   \\
72.855 & $-$1063.7741& 421.0850& 11194& 0.35& 11100\\
73.031 &  0.0000   & 0.0000  & 11744& 0.49& 11978\\
73.119 & $-$39.3480  &$-$27.3450 & 1431 & 0.58& 1613 \\
73.821 & $-$1163.3500& 155.9080& 1134 & 0.19& 1149 \\
74.173 & $-$48.5034  &$-$43.3810 & 2845 & 0.39& 3600 \\
74.436 & $-$43.6441  &$-$45.4110 & 6641 & 0.54& 5863 \\
75.139 & $-$9.8909   & 12.9190 & 371  & 0.35& 357  \\
75.579 & $-$46.4093  &$-$46.6400 & 316  & 0.36& 302 \\
75.754 &  7.3175   & 14.7340 & 431  & 0.28& 431 \\
76.106 &  0.1992   & 16.2260 & 3900 & 0.27& 4007\\
81.289 & $-$1122.9459& 420.7460& 1619 & 0.78& 1373\\
81.553 & $-$1100.4663& 436.9820& 447  & 0.77& 386 \\
81.816 & $-$1103.1221& 438.3080& 513  & 0.27& 515 \\
81.904 & $-$1116.3250& 407.0410& 1010 & 0.31& 1103\\
82.168 & $-$1113.5523& 420.1690& 5062 & 0.32& 5183\\
83.925 & $-$1116.4418& 417.6580& 1652 & 0.66& 1816\\
\multicolumn{6}{l}{G36.705$+$00.096}\\         
52.560 &  1.3392 &  1.1530 & 348 & 0.33& 353   \\
53.087 &  0.0000 &  0.0000 & 7576& 0.31& 7608  \\
53.790 & $-$0.9167 & $-$0.1560 & 2413& 0.48& 2470  \\
54.405 & $-$1.8934 & $-$1.2010 & 4629& 0.65& 4404  \\
54.844 & $-$2.0102 & $-$9.9610 & 1309& 0.43& 1343 \\
55.108 & $-$1.5579 & $-$3.2250 & 420 & $-$   & $-$  \\
62.137 & $-$14.1016&  52.0590& 1974& 0.45& 1969 \\
63.015 & $-$0.3940 &  45.6080& 153 & 0.47& 154  \\
\multicolumn{6}{l}{G37.030$-$00.039}\\
78.585 & 0.0000 &    0.0000&  691&  0.47& 656 \\
78.761 &$-$0.4374 &    0.2050&  224&  $-$   &  $-$\\
\hline
\end{supertabular}
\end{table}
\begin{table}
\centering
\addtocounter{table}{-1}
\caption{ continued.}
\begin{supertabular}{rrrccc}
\hline\hline       
\\
V$_{\rm LSR}$ & $\Delta$RA & $\Delta$Dec & S$_{\rm p}$& FWHM & S$_{\rm amp}$\\
(km\,s$^{-1}$) & (mas)     &  (mas)      & (mJy\,beam$^{-1}$)&
(km\,s$^{-1}$)& (mJy\,beam$^{-1}$)\\
\hline
\\
\multicolumn{6}{l}{G37.598$+$00.425}\\       
82.802 & 91.6302 &  $-$11.1510& 251 & $-$   & $-$ \\
85.789 & 0.0000  &   0.0000 & 3911& 0.43& 3826\\
85.438 &$-$0.3322  &  $-$0.8860 & 891 & 0.33& 902 \\
86.931 & 26.0379 &   12.9580& 2407& 0.32& 2482\\
87.195 & 25.4768 &   13.1770& 465 & $-$   & $-$ \\
87.283 & 39.3823 &   6.436  & 2543& 0.45& 2661\\
88.600 & 63.5139 &   4.0650 & 395 & $-$   &  $-$ \\
88.688 & 66.2298 &  $-$7.5300 & 202 & $-$   &  $-$   \\
\multicolumn{6}{l}{G38.038$-$00.300}\\        
55.656&  0.0000  &   0.0000 & 2168& 0.34& 1996\\
57.062&  6.2921  &  $-$15.5470& 507 & $-$   &  $-$\\
57.413&  9.3731  &  $-$16.4000& 1046& 0.38& 980 \\
58.204&  13.9385 &  $-$1.8140 & 1434& 0.25& 1444\\
59.522&  20.8441 &  $-$15.0610& 215 & $-$   &  $-$\\
59.522&  19.9328 &   12.8000& 224 & $-$   &  $-$\\
59.609&  19.0545 &  $-$3.8880 & 281 & $-$   &$-$\\
\multicolumn{6}{l}{G38.203$-$00.067}\\
79.640&  0.0000   &  0.0000  & 828& 0.28& 707 \\
80.518& $-$54.8141  &  20.5190 & 169& $-$   &   $-$  \\
80.606&  11.2600  &  13.4970 & 157& 0.38& 160 \\
83.856& $-$101.8398 &  154.2200& 173& $-$   &$-$     \\
83.944& $-$35.3736  &  29.9400 & 152& $-$   &   $-$  \\
83.944& $-$35.2315  &  44.9650 & 195& $-$   &$-$     \\
84.208& $-$33.8743  &  31.6240 & 198& $-$   &   $-$  \\
\multicolumn{6}{l}{G39.100$+$00.491}\\
14.630 &$-$5.7037   & $-$38.5700 & 222& $-$   &  $-$   \\
14.630 &$-$7.8346   & $-$11.4310 & 453& 0.28& 446 \\
15.772 & 13.3126  & $-$22.0650 & 314& $-$   &   $-$  \\
15.860 & 115.0769 &  103.8770& 623& $-$   &   $-$  \\
15.860 & 116.6086 &  75.8880 & 545& $-$   &   $-$ \\
15.860 & 13.3112  & $-$22.0040 & 769& $-$   & $-$ \\
15.245 & 0.0000   &  0.0000  &2068& 0.37& 1803\\
15.684 &$-$5.1254   & $-$5.8070  &1140& 0.62& 1158\\
15.948 &$-$4.9192   & $-$7.1320  &1042& 0.23& 1108\\
17.617 & 6.6630   & $-$8.1800  & 696& 0.32& 654 \\
\\
\hline\hline                                   
\end{supertabular}                                                                                
\end{table}

\end{document}